\pdfoutput=1

\documentclass[11pt,twoside,a4paper,cmspaper,final,collab]{cms-tdr}
\def\svnVersion{452199P}\def\cmsCernNoTag{CERN-EP-2017-258}\def\cmsCernDate{\today}\def\cmsMessage{Published in the Journal of High Energy Physics as \href{http://dx.doi.org/10.1007/JHEP03(2018)115}{\doi{10.1007/JHEP03(2018)115.}}}

\begin{document}\cmsNoteHeader{TOP-16-023}

\hyphenation{had-ron-i-za-tion}
\hyphenation{cal-or-i-me-ter}
\hyphenation{de-vices}
\RCS$Revision: 445016 $
\RCS$HeadURL: svn+ssh://svn.cern.ch/reps/tdr2/papers/TOP-16-023/trunk/TOP-16-023.tex $
\RCS$Id: TOP-16-023.tex 445016 2018-02-08 22:33:46Z alverson $

\newcommand{\stt}{\ensuremath{\sigma_{\ttbar}}\xspace}
\newcommand{\sqrts}{\ensuremath{\sqrt{s}}\xspace}
\providecommand{\empm}{\ensuremath{\Pe^\pm \Pgm^\mp}\xspace}
\providecommand{\mmpm}{\ensuremath{\Pgm^\pm \Pgm^\mp}\xspace}
\providecommand{\mll}{\ensuremath{M_{\ell\ell}}\xspace}
\providecommand{\WV}{\ensuremath{\PW\cmsSymbolFace{V}}\xspace}
\providecommand{\nonW}{{non-\PW/\cPZ{}}\xspace}
\providecommand{\minDeltaR}{\ensuremath{\Delta R_\text{min}(j,j')}\xspace}
\newlength{\cmsTabSkip}\setlength{\cmsTabSkip}{2ex}

\cmsNoteHeader{TOP-16-023}
\title{Measurement of the inclusive \ttbar cross section  in pp collisions at $\sqrt{s}$ = 5.02\TeV using final states with at least one charged lepton}

\date{\today}

\abstract{
The top quark pair production cross section (\stt) is measured
for the first time in pp collisions at a center-of-mass energy of 5.02\TeV. The data were collected by the CMS experiment at the LHC and correspond to an integrated luminosity of 27.4\pbinv. The measurement is performed by analyzing events with at least one charged lepton. The measured cross section is
$ \stt  =  69.5 \pm 6.1\stat\pm 5.6\syst\pm 1.6\lum\unit{pb}$, with a total relative uncertainty of 12\%.
The result is in agreement with the expectation from the standard model.
The impact of the presented measurement on the determination of the gluon distribution function is investigated.
}

\hypersetup{
pdfauthor={CMS Collaboration},
pdftitle={Measurement of the inclusive ttbar cross section in pp collisions at sqrt(s) = 5.02 TeV using final states with at least one charged lepton},
pdfsubject={CMS},
pdfkeywords={CMS, physics, top}}

\maketitle

\section{Introduction}
\label{sec:intr}
The top quark, the heaviest elementary particle in the standard model (SM), has been the subject of numerous detailed studies using hadron-hadron collisions.
The pair production (\ttbar) cross section (\stt) as a function of center-of-mass energy
can be of interest for the extraction of the top quark pole mass~\cite{polemass} and has been used to constrain the gluon distribution function~\cite{Sirunyan:2017azo} at
large fractions $x$ of the proton longitudinal momentum carried by the gluon, where the gluon distribution is poorly known.
 Precise measurements of \stt in proton-proton (pp) collisions have been published at \sqrts values of 7 and 8~\cite{ATLAStt12,ATLAStt6,Khachatryan:2016mqs,Khachatryan:2016yzq} and 13\TeV~\cite{ATLAStt13TeV2,Khachatryan:2015uqb,CMS-PAS-TOP-16-005,CMS-PAS-TOP-16-006} by the ATLAS and CMS Collaborations at the LHC.

 In November 2015, the LHC delivered pp collisions at $\sqrts = 5.02$\TeV.
 The fraction of \ttbar events initiated by gluon-gluon collisions grows monotonically with \sqrts. It is around 73\% at 5.02\TeV,
as calculated with \POWHEG~(v2)~\cite{powheg,powheg2,powheghvq} at next-to-leading order (NLO) using the NNPDF3.0 NLO~\cite{Ball:2014uwa} parton distribution functions (PDFs),
and increases to around 86\% at 13\TeV, making this new data set partially complementary to the higher-energy samples.
Measurements of \ttbar production at various \sqrts probe different values of $x$ and thus can provide complementary information on the gluon distribution.
In addition, future measurements of \stt in nuclear collisions at the same nucleon-nucleon center-of-mass energy~\cite{dEnterria:2015jna,CMS-PAS-FTR-13-025}
would profit from the availability of a reference measurement in pp collisions at $\sqrts = 5.02\TeV$, without the need to extrapolate from measurements at different \sqrts.
This has already been demonstrated with the first observation of the \ttbar process using proton-nucleus collisions at a higher nucleon-nucleon center-of-mass energy~\cite{ttbarpPb}.

In the SM, top quarks in $\Pp\Pp$ collisions are mostly produced as \ttbar pairs. Each top quark
decays predominantly to a \PW{} boson and a bottom (b) quark. The \ttbar events are categorized according to the decay of the two W bosons.
In \ttbar events
where one \PW{} boson decays leptonically and the other hadronically ($\ell$+jets channel), the final state
presents a typical signature of one isolated lepton, missing transverse momentum, two jets from the W boson hadronic decay, and two
jets coming from the hadronization of the b quarks (``\PQb\ jets'').
On the other hand, in \ttbar events where both \PW{} bosons decay leptonically (dilepton channel), the final state contains two leptons of opposite electric charge, missing transverse momentum, and at least two b jets.
The $\ell$+jets channel has a large branching ratio with a moderate amount of background, while the dilepton channel is
characterized by a high purity.

This analysis represents the first measurement of \stt in pp collisions at $\sqrts = 5.02$\TeV using \ttbar candidate events
with $\ell$+jets, where leptons are either electrons ($\ell=\Pe$) or muons ($\ell=\mu$), and dilepton (\empm or \mmpm) final states. In the former case,
\stt is extracted by a fit to the distribution of a kinematic variable for different categories of lepton flavor and jet multiplicity, while in the latter an event counting  approach is used.
The two results are then combined in the final measurement, which is used as input to a quantum chromodynamics (QCD) analysis at next-to-next-to-leading
order (NNLO) to investigate the impact on the determination of the gluon distribution in the less-explored kinematic range of $x \gtrsim 0.1$.

This paper is structured as follows. Section~\ref{sec:cms} describes the CMS detector.
Section~\ref{sec:samples} gives a
summary of the data and simulated samples used. After the discussion of the object reconstruction in Section~\ref{sec:obj},
and of the trigger and event selection in Section~\ref{sec:evtSel},
Section~\ref{sec:bkg} describes the determination of the background sources. The systematic uncertainties
are discussed in Section~\ref{sec:syst}. The extraction of \stt is presented in Section~\ref{sec:res} and the impact
of the presented measurement on the determination of the proton PDFs is discussed in Section~\ref{sec:pdfs}. A summary of all the results is given in Section~\ref{sec:summ}.

\section{The CMS detector}
\label{sec:cms}
The central feature of the CMS apparatus is a superconducting solenoid of 6\unit{m} internal diameter, providing a
magnetic field of 3.8\unit{T} parallel to the beam direction.

Within the solenoid volume are a silicon pixel and strip tracker, a lead tungstate crystal electromagnetic calorimeter (ECAL), and a brass and scintillator hadron calorimeter (HCAL), each composed of a barrel and two endcap sections.
A preshower detector, consisting of two planes of silicon sensors interleaved with about 3 radiation lengths of lead,
is located in front of the endcap regions of ECAL.
Hadron forward calorimeters using steel as an absorber and quartz fibers as the sensitive material
extend the pseudorapidity coverage provided by the barrel and endcap detectors from $\abs{\eta} = 3.0$ to $5.2$.

Charged particle trajectories with $\abs{\eta} < 2.5$ are measured by the tracker system, while the energy deposits in ECAL
and HCAL cells are summed to define the calorimeter tower energies, subsequently used to calculate the energies and directions of hadronic jets.
Muons are detected in the pseudorapidity window $\abs{\eta} < 2.4$ in gas-ionization detectors embedded in the steel flux-return yoke outside the solenoid.
Photons and electrons are reconstructed by their deposited energy in groups of ECAL crystals (``clusters'').
Events of interest are selected using a two-tiered trigger system~\cite{hlt}. The first level, composed of custom hardware processors, uses information from the calorimeters and muon detectors to select events at a rate of around 100\unit{kHz} within a time interval of less than 4\mus. The second level, known as the high-level trigger, consists of a farm of processors running a version of the full event reconstruction software optimized for fast processing, and reduces the event rate to around 1\unit{kHz} before data storage.

A more detailed description of the CMS detector, together with a definition of the coordinate system used and the  relevant kinematic variables, can be found in Ref.~\cite{Chatrchyan:2008zzk}.
\section{Data, simulated samples and theoretical cross section}
\label{sec:samples}

This analysis is based on an integrated luminosity of $27.4 \pm 0.6\pbinv$~\cite{CMS-PAS-LUM-16-001}.
The presence of multiple proton collisions in the same or nearby bunch crossings (``pileup'') results in an average number of overlapping interactions estimated online to be 1.4, assuming a total inelastic cross section of 65\unit{mb}.

Several Monte Carlo (MC) event generators are used to simulate signal and background events. The NLO \POWHEG~(v2)~\cite{powheg,powheg2,powheghvq}
generator is used for \ttbar events, assuming a value of 172.5\GeV for the top quark mass ($m_\text{top}$).
These events are passed to \PYTHIA~(v8.205)~\cite{Sjostrand:2006za,Sjostrand:2014zea}
to simulate parton showering, hadronization, and  the underlying event, using the CUETP8M1~\cite{Khachatryan:2015pea,Skands:2014pea} tune for the default \ttbar MC sample.
The NNPDF3.0 NLO PDFs with strong coupling $\alpha_\mathrm{s}(M_{\Z})=0.118$ at the Z boson mass scale $M_{\Z}$ are utilized in the MC calculations.

The {\MGvATNLO~(v5\_2.2.2) generator~\cite{amcatnlo} is used to simulate W boson production with additional jets (\PW+jets), and high-mass ($>$50\GeV) Drell--Yan quark-antiquark
annihilation into lepton-antilepton pairs
through Z boson or virtual-photon exchange (referred to as ``Z$/\gamma^{*}$'').
The simulation includes up to two extra partons at matrix element level, and the \textsc{FxFx} merging
procedure~\cite{fxfx} is used to interface with \PYTHIA. Low-mass Z/$\gamma^{*}$ events (20--50\GeV) are simulated with \PYTHIA.
The normalization of the \PW+jets and Z/$\gamma^{*}$ processes is either derived from data (in the dilepton channel) or estimated based on the NNLO cross sections (in the $\ell$+jets channel) from the \FEWZ program (v3.1.b2)~\cite{fewz}.
Single top quark plus \PW{} boson events ($\PQt\PW$) are simulated using \POWHEG~(v1)~\cite{powheg1,powheg3} interfaced with \PYTHIA,
and are normalized to the approximate NNLO cross sections~\cite{Kidonakis:2013zqa}.
The contributions from \PW\PW\ and \PW\cPZ\ production (referred to as ``WV'') are simulated with \PYTHIA, and are normalized
to the NLO cross sections calculated with the \MCFM~(v8.0) program~\cite{mcfm}.
All generated events undergo a full \GEANTfour~\cite{geant} simulation of the detector response.

The expected signal yields are normalized to the value of the SM prediction for the \ttbar production cross section:
\begin{equation}
\label{eq:theory}
\sigma_{\tt}^{\mathrm{NNLO}} = 68.9\ ^{+1.9}_{-2.3}\,\text{(scale)} \pm 2.3\,\mathrm{(PDF)}\ ^{+1.4}_{-1.0}\ (\alpha_\mathrm{s})\unit{pb},
\end{equation}
as calculated with the \textsc{Top++} program~\cite{top++} at NNLO in perturbative QCD, including soft-gluon resummation
at next-to-next-to-leading-logarithmic order~\cite{mitov}, using the NNPDF 3.0 NNLO PDF set, with
$\alpha_\mathrm{s}(M_{\Z})=0.118$ and $m_\text{top} = 172.5\GeV$.
The systematic uncertainties in the theoretical \ttbar cross section are associated with the choice of the renormalization ($\mu_{\mathrm{R}}$) and factorization ($\mu_{\mathrm{F}}$) scales--nominally set at $\mu_{\mathrm{R}}=\mu_{\mathrm{F}}=\sqrt{\smash[b]{m_\text{top}^{2}+p_{\mathrm{T},\text{top}}^{2}}}$ with $p_{\mathrm{T},\text{top}}$ the top quark transverse momentum--as well as with the PDF set and the $\alpha_s$ value.
The uncertainty of 0.1\% in the LHC beam energy~\cite{Wenninger:2254678} translates into an additional uncertainty of 0.22\unit{pb} in the expected cross section, with negligible impact on the acceptance of any of the channels included in this analysis.

\section{Object reconstruction}
\label{sec:obj}

The particle-flow (PF) algorithm~\cite{PFPAS1} is used to reconstruct and identify individual particles using an optimized combination of information from the various elements of the CMS detector.

The electron momentum is calculated by combining the energy measurement in the ECAL with the momentum measurement
in the tracker, taking into account the bremsstrahlung photons spatially compatible with originating from the electron track.
The momentum resolution for electrons with transverse momentum $\pt \approx 45\GeV$ from $\Z \to \Pe \Pe$
decays ranges from 1.7\% for nonshowering electrons in the barrel region to 4.5\% for showering electrons in
the endcaps~\cite{Khachatryan:2015hwa}.
Muon candidates are reconstructed from a combination of the information collected by the muon spectrometer and the silicon tracker.
This results in a relative \pt resolution of
1.3--2.0\% in the barrel and better than 6\% in the endcaps, for muons with $20 <\pt < 100\GeV$
and within the range $\abs{\eta} < 2.4$~\cite{Chatrchyan:2012xi, muid}.
The photon energy is directly obtained from the ECAL measurement, corrected for zero-suppression effects.
The charged hadron energies are determined from a combination of their momenta measured in the tracker and the matching ECAL and HCAL energy deposits, corrected for zero-suppression effects and for the response function of the calorimeters to hadronic showers. Finally, the neutral hadron energies are obtained from the corresponding corrected ECAL and HCAL energies.

The missing transverse momentum vector is defined as the negative vector sum of the momenta of all reconstructed PF candidates in an event,
projected onto the plane perpendicular to the direction of the proton beams.
Its magnitude is referred to as \ptmiss and the corrections to jet momenta are propagated to the \ptmiss calculation~\cite{CMS-PAS-JME-16-004}.

The reconstructed vertex with the largest value of summed physics-object $\pt^2$ is taken to be the primary $\Pp\Pp$ interaction vertex.
The physics objects are the jets, clustered using the jet finding algorithm~\cite{antikt,Cacciari:2011ma} with the tracks assigned to the vertex as inputs, and the associated \ptmiss.
The isolation of electron and muon candidates from nearby jet activity is then evaluated as follows.
For electron and muon candidates, a cone of $\Delta R = 0.3$ and 0.4, respectively, is constructed around the direction of the lepton track at the primary event vertex,
where $\Delta R$ is defined as $\sqrt{\smash[b]{(\Delta \eta)^2 + (\Delta \phi)^2}}$, and $\Delta\eta$ and $\Delta \phi$ are the differences in pseudorapidity and
azimuthal angle between the directions of the lepton and another particle.
A relative isolation discriminant, $I_\text{rel}$, is calculated by the ratio between the scalar \pt sum of all particle candidates inside the cone consistent with originating from the primary vertex and the \pt of the lepton candidate. In this sum, we exclude the \pt of the lepton candidate.
The neutral particle contribution to $I_\text{rel}$ is corrected for energy deposits from pileup interactions using different techniques for electrons and muons. For muons, half of the total \pt of the charged hadron PF candidates not originating from the primary vertex is subtracted. The factor of one half accounts for the different fraction of charged and neutral particles in the cone. For electrons, the \textsc{FastJet} technique~\cite{Cacciari:2007fd} is used, in which the median of the energy-density distribution of neutral particles (within the area of any jet in the event) multiplied by the geometric area of the isolation cone--scaled by a factor that accounts for the residual $\eta$-dependence of the average energy deposition due to pileup--is subtracted.

The efficiency of the lepton selection is measured using a ``tag-and-probe'' method in same-flavor dilepton events enriched in \Z
boson candidates, following the method of Ref.~\cite{inclusWZ3pb}.
 The sample of $\Z\to\Pgmp\Pgmm$ events used for muon efficiency extraction is selected by the same trigger requirement used by the main analysis (Section~\ref{sec:evtSel}).
 The $\Z\to \Pep\Pem$ sample for electron efficiency extraction makes use of events that satisfy a diphoton trigger with symmetric transverse energy, $\ET=\sum_{i} E_i \sin\theta_i$, thresholds of $\ET=15\GeV$ covering the full tracker acceptance, where $E_i$ is the energy seen by the calorimeters for the $i$th particle, $\theta_i$ is the polar angle of particle $i$, and the sum is over all particles emitted into a fixed solid angle in the event. Pairs of photon candidates above the \ET threshold are accepted only if their invariant mass is above 50\GeV. The trigger selection requires a loose identification using cluster shower shapes and a selection based on the ratio of the hadronic to the electromagnetic energy of the photon candidates.
 Based on a comparison of the lepton
selection efficiency in data and simulation, the event yield in
simulation is corrected using data-to-simulation scale factors.

Jets are reconstructed from the PF candidates using the anti-\kt clustering algorithm~\cite{antikt} with a
distance parameter of 0.4.
Jets closer than
$\Delta R = 0.3$ to the nearest muon or electron are discarded. Jet energy corrections extracted from full detector
simulation are also applied as a function of jet $\pt$ and $\eta$~\cite{JESPUB} to data and simulation. A residual
correction to the data is applied to account for the discrepancy between data and simulation in the jet response.

\section{Event selection}
\label{sec:evtSel}

The event sample is selected by a loose online trigger
 and further filtered
offline to remove noncollision events, such as beam-gas interactions or cosmic rays.
Collision events containing one high-$\pt$ electron (muon) candidate are selected online by requiring
values of  \et (\pt) greater than 40 (15)\GeV and of $\abs{\eta}$ less than 3.1 (2.5).
The measured trigger efficiency for each decay channel, relative to the final selection, is higher than 90\%.

In the $\ell$+jets analysis, electron candidates are selected if they have $\pt>40\GeV$ and $\abs{\eta}<2.1$.
Further identification and isolation criteria are applied to the electron candidates.
Electrons reconstructed in the ECAL barrel (endcap) are required to have $I_\text{rel}< 4$ ($5$)$\%$.
Electron candidates in the  $1.44<\abs{\eta}<1.57$ region, i.e., in the transition region between the barrel and endcap sections of the ECAL, are excluded because the
reconstruction of an electron object in this region is less efficient.
Muons are required  to have $\pt>25\GeV$ and $\abs{\eta}<2.1$.
Additional identification criteria are applied and $I_\text{rel}$ is required to be $<15\%$.
 Events are rejected if they contain extra electrons or muons identified using a looser set of identification criteria and have $\pt>10$ or $15\GeV$, respectively.

The distinct signature of two \PQb jets, expected in \ttbar decays, is rare in background events, and thus is exploited in the $\ell$+jets analysis.
Backgrounds from \PW+jets, QCD multijet, and Z/$\gamma^{*}$ events are controlled by counting the number of b jets in the selected events.
In addition, two light-flavor jets are expected to be produced in the decay of one of the \PW{} bosons for signal events.
The correlation in phase space of these light jets carries a distinctive hallmark with respect to the main backgrounds.
To that end, jets are selected if they have $\pt>30\GeV$ and $\abs{\eta}<2.4$.
The flavor of the jets is identified using a combined secondary vertex algorithm~\cite{btaggingRun2}
with an operating point that yields a b jet identification efficiency of about 70\%, and misidentification (mistag)
probabilities of about 1\% and 15\% for light-flavor (\PQu, \PQd, \PQs,
and gluons) and \cPqc\ jets, respectively.
The event selection requires at least two non-\PQb-tagged jets to be identified as candidates from the W boson hadronic decay.
Additional jets passing the \PQb quark identification criteria are counted and used to classify the selected events
in none (0 \PQb), exactly one (1 \PQb), or at least two ($\geq$2 \PQb) tagged jet categories. The efficiency of the \PQb jet identification algorithm is measured {\it in situ},
simultaneously with the signal cross section.

Dilepton events are required to contain at least one muon candidate at trigger level.
No requirement on the presence of electron candidates is made at trigger level owing to the relatively high-$\et$ threshold (40\GeV) of the trigger.
Electrons are selected if they have $\pt > 20\GeV$, $\abs{\eta} < 2.4$, and $I_\text{rel}<9$ (or 12)\% if
in the barrel (or one of the endcaps). As in the $\ell$+jets channel, electrons detected
in the transition region between the barrel and endcap sections of the ECAL are excluded.
Muons are required to have $\pt > 18\GeV$, $\abs{\eta} < 2.1$, and $I_\text{rel}<15\%$.
At least two jets satisfying the criteria  $\pt> 25\GeV$ and $\abs{\eta}< 3$ are required.
Events are subsequently selected if they have
a pair of leptons with opposite charge (\empm or \mmpm) passing the requirements listed above. In events with more than
one pair of leptons passing the above selection, the two leptons
of opposite charge that yield the highest scalar \pt sum are selected.

Candidate events with dilepton invariant masses of $\mll <20$\GeV are removed to suppress events from decays of heavy-flavor resonances and low-mass Z/$\gamma^{*}$ processes.
Dilepton events with two muons in the final state are still dominated by the Z/$\gamma^{*}$ background. In order to suppress
this contribution, events in the Z boson mass window of $76<\mll <106$\GeV are vetoed in this channel.
To further suppress the Z/$\gamma^{*}$ events, a requirement on \ptmiss of $>$35\GeV is imposed.

In both the $\ell$+jets and dilepton analyses, events with \Pgt\ leptons are considered as signal if they decay to electrons or muons that satisfy the selection requirements, and are included in the simulation.
\section{Background estimation}
\label{sec:bkg}

\subsection{The \texorpdfstring{$\ell$}{l}+jets final state}
\label{sec:bkg_ljets}

In the $\ell$+jets analysis, the contributions of all background processes are estimated from simulation,
with the exception of the QCD multijet background. Due to its large cross section,
there is a nonnegligible contribution from the latter faking a \ttbar event with $\ell$+jets in the final state.
Both the contribution from hard fragmentation of c and b quarks whose hadrons decay semileptonically,
and the contribution from misidentified leptons, such as from either punch-through hadrons or collimated jets with a high electromagnetic fraction,
can yield $\ell$+jets-like topologies.

The estimation of the QCD multijet background is separately performed for the events with 0, 1, or $\geq$2 \PQb{} jets
using a control region where either the muon candidate fails a looser isolation requirement ($I_\text{rel}<20\%$)
or the electron candidate fails the identification criteria.
The choice of the QCD multijet control region has been made in such a way as to minimize the contamination due to the signal and \PW+jets events, while retaining a large number of events in the sample for the estimation of this type of background.
The initial normalization of the QCD multijet contribution in the signal region is derived from events with
$\ptmiss<20\GeV$ (``reduced-signal'' region). Events in both the reduced-signal and control regions fulfilling this requirement
are counted.
After subtracting the expected contributions from non-QCD processes,
the ratio between the numbers of events observed in the reduced-signal region and in the control region,
is used as a transfer factor to normalize the QCD multijet background estimate.
In both the electron and muon channels, a 30\% uncertainty is assigned to
the estimate of the expected contribution from non-QCD processes,
estimated after varying the QCD scales in the \PW+jets simulation.
This uncertainty propagates as both a normalization and a shape uncertainty in the predicted distributions for the QCD multijet processes.
The variations are applied independently in the reduced-signal and control regions in order to determine an uncertainty envelope.
A more accurate normalization for this contribution is obtained by the fit performed to extract the final cross section, described in Section~\ref{sec:res_ljets}.

\subsection{The dilepton final state}
\label{sec:bkg_dil}

Final states with two genuine leptons can originate from background processes,
primarily from Z/$\gamma^{*} \to \tau^+\tau^-$ (where the $\tau$ leptonic decays can yield $\empm$ or $\mmpm$ plus \ptmiss due to the neutrinos), $\PQt\PW$, and \WV events. Other background
sources, such as \PW+jets events or \ttbar production in the $\ell$+jets final state, can
contaminate the signal sample if a jet is misidentified as a lepton, or if an event contains a
lepton from the decay of b or c hadrons. These are included in the ``\nonW''
category, since genuine leptons are defined as originating from decays of W or Z bosons.
The yields
from $\PQt\PW$ and \WV events are estimated from simulation, while the contribution of the Z/$\gamma^{*}$ background is evaluated using control samples in data.
The rate of \nonW backgrounds is extracted from control samples in data for the \empm channel and is estimated from simulation for the \mmpm channel.

A scale factor for the Z/$\gamma^{*}$ background normalization is estimated, as in Ref.~\cite{Khachatryan:2010ez},
from the number of events within the \Z boson mass window in data, which is extrapolated to
the number of events outside the window.
A scale factor of
$0.91 \pm 0.14\stat$ is obtained in the \empm channel, and $0.96 \pm 0.78\stat$ in the \mmpm channel.
The estimation is performed using events with at least two jets, and the dependence on different jet multiplicities is discussed in Section~\ref{sec:syst}.

The \nonW background in the \empm channel is estimated using an extrapolation from a control region of same-sign (SS) dilepton
events to the signal region of opposite-sign (OS) dileptons.
The SS control region is defined using the same criteria as for the
nominal signal region, except requiring dilepton pairs of the same
charge. The muon isolation requirement is relaxed in order to enhance the number of events.
The SS dilepton events predominantly contain at least one
misidentified lepton.
Other SM processes produce genuine SS or charge-misidentified dilepton events
with significantly smaller rates; these are estimated using simulation and subtracted from the
observed number of events in data.
The scaling from the SS control region in data to the signal region is
performed using an extrapolation factor extracted from MC simulation,
given by the ratio of the number of OS events with misidentified
leptons to the number of SS events with misidentified leptons.
The resulting estimate for the non-W/Z background is
$1.0 \pm 0.9\stat$ events, where the central value comes from the estimation using events with at least two jets. No particular dependence of this scale factor is observed for different jet multiplicities within the large statistical uncertainty.
\section{Systematic uncertainties}
\label{sec:syst}

The integrated luminosity has been estimated offline using a pixel cluster counting method~\cite{CMS-PAS-LUM-16-001}. The estimation takes into account normalization uncertainties
and uncertainties related to the different conditions during typical physics periods relative to the specially tailored beam-separation scans, adding up to a total uncertainty of $\pm2.3\%$.

The uncertainties in the electron trigger efficiency (1.5\%) and the identification and isolation efficiency (2.5\%) are estimated by changing the values of the
data-to-simulation scale factors within their uncertainties,
as obtained from the ``tag-and-probe'' method.
The uncertainty in the muon identification and isolation efficiency, including the trigger efficiency, is 3\% and covers one standard deviation of the scale factor from unity.

The impact of the uncertainty in the jet energy scale (JES) is estimated by changing the
\pt- and $\eta$-dependent JES corrections by a constant 2.8\%~\cite{JESPUB,JES8TeVPUB}.
The uncertainty in jet energy resolution (JER)
is estimated through $\eta$-dependent
changes in the JER corrections to the simulation~\cite{JESPUB,JES8TeVPUB}.
The uncertainty arising from the use of \ptmiss in the \mmpm channel is dominated by the unclustered energy contribution to \ptmiss~\cite{CMS-PAS-JME-16-004}.
Finally, a 30\% uncertainty is conservatively assigned to the jet misidentification probability in the $\ell$+jets analysis,
as no dedicated measurement of this quantity has been performed for the considered data set.

Theoretical uncertainties in the simulation of \ttbar production cause a systematic bias
related to the missing higher-order diagrams in \POWHEG, which is
estimated through studies of the signal modeling by modifying the $\mu_\mathrm{R},\mu_\mathrm{F}$ scales within a factor of two with respect to their nominal value.
In the $\ell$+jets analysis, the impact of the $\mu_\mathrm{R},\mu_\mathrm{F}$ variations are examined independently, while in the dilepton analysis they are varied simultaneously.
In both analyses, these variations are applied independently at the matrix element (ME) and parton shower (PS) levels.
 The uncertainty arising from the hadronization model mainly affects the
JES and the fragmentation of jets.
 The hadronization uncertainty is determined by comparing samples of
events generated with \POWHEG, where the hadronization is either
modeled with \PYTHIA or \HERWIGpp~(v2.7.1)~\cite{herwigpp}. This also
accounts for differences in the PS model and the underlying event.
 The uncertainty from the choice of PDF is determined by
reweighting the sample of simulated  \ttbar events according to the
root-mean-square (RMS) variation of the NNPDF3.0 replica set.
Two extra variations of $\alpha_\mathrm{s}$ are added in quadrature
to determine the total PDF uncertainty.

In the $\ell$+jets analysis, the uncertainty in the choice of the $\mu_\mathrm{R},\mu_\mathrm{F}$ scales in the \PW+jets simulation is taken into account
by considering alternative shapes and yields after varying independently the $\mu_\mathrm{R},\mu_\mathrm{F}$ scales,
following a similar procedure to that described above for the signal.
Due to the finite event count in the \PW+jets simulated sample, an additional bin-by-bin uncertainty is assigned by generating an alternative shape to fit (see Section~\ref{sec:res_ljets}),
where the bin prediction is varied by $\pm$1 standard deviation, while keeping all the other bins at their nominal expectation.
The uncertainty assigned to the QCD multijet background includes the statistical uncertainty in the data, and
the uncertainty from the non-QCD multijet contributions subtracted from the control region, as described in Section~\ref{sec:bkg_ljets},
and an additional 30\%--100\% normalization uncertainty.
The latter depends on the event category and stems from the measured difference with respect to an alternative estimate of the QCD normalization
based on the transverse mass, $m_\text{T}$, of the lepton and \ptmiss system. The magnitude of $m_\text{T}$ equals $\sqrt{\smash[b]{2  \pt  \ptmiss  (1-\cos \Delta \phi)}}$,
where \pt is the lepton transverse momentum and $\Delta\phi$ is the azimuthal angle between the lepton and the direction of \ptvecmiss.
Finally, a 30\% normalization uncertainty in the theoretical tW, Z/$\gamma^{*}$, and WV background cross sections is assigned~\cite{Khachatryan:2016mqs},
given the previously unexplored $\sqrt{s}$  value and that the final states contain several jets.

In the dilepton channel, an uncertainty of 30\% is assumed~\cite{Khachatryan:2016mqs} for the cross sections of the $\PQt\PW$ and \WV\ backgrounds to cover the theoretical uncertainties
and the effect of finite simulated samples. The uncertainty in the Z/$\gamma^{*}$ estimation is calculated by combining in quadrature the statistical uncertainty and
an additional 30\% from the variation of the scale factor in the different levels of selection, resulting in uncertainties of about 30 and 80\% in the \empm and \mmpm channels,
respectively.
The systematic uncertainty in the non-W/Z background is estimated to be 90\% in the \empm channel and is dominated by the statistical uncertainty in the method.
Owing to the limited sample size in the data, the method cannot be applied in the \mmpm channel. The estimation is therefore based on MC simulation, and an uncertainty of 100\% is conservatively assigned.
\section{Measurement of the \texorpdfstring{\ttbar}{ttbar} cross section}
\label{sec:res}

\subsection{The \texorpdfstring{$\ell$}{l}+jets final state}
\label{sec:res_ljets}

In the $\ell$+jets analysis, the \ttbar cross section is measured in a fiducial phase space by means of a fit.
Two variables were independently considered for the fit, which are sensitive to the resonant behavior of the light jets produced from the \PW{} boson hadronic decay in a \ttbar event.
Given that these light jets, here denoted by $j$ and $j'$, are correlated during production, they
are also expected to be closer in phase space when compared to pairs of other jets in the event.
The angular distance $\Delta R$ can thus be used as a metric to rank all pairs of non-\PQb-tagged jets in the event,
maximizing the probability of selecting those from the \PW{} boson hadronic decay in cases where more than two non-\PQb-tagged jets are found.
From simulation we expect that the signal peaks at low $\Delta R$, while the background is uniformly distributed up to
$\Delta R\approx 3$. Above that value, fewer events are expected and background processes are predicted to dominate.
The invariant mass $M(j,j')$ of jets $j$ and $j'$ also has a distinctive peaking feature for the signal in contrast with a smooth background continuum.
From simulation we expect that the minimum angular distance $\Delta R$ between all pairs of jets $j$ and $j'$, \minDeltaR,
is robust against signal modeling uncertainties
such as the choice of the $\mu_\mathrm{R},\mu_\mathrm{F}$ scales and jet energy scale and resolution,
while the $M(j,j')$ variable tends to be more affected by such uncertainties.
Owing to its more robust systematic uncertainties and signal-to-background discrimination power, the \minDeltaR
variable is used to extract the \ttbar cross section.

In order to maximize the sensitivity of the analysis, the \minDeltaR distributions are categorized
according to the number of jets--in addition to the ones assigned to the W boson hadronic decay--passing the b quark identification criteria. In total, 6 categories are used, corresponding to electron or muon events with
0, 1, or $\geq$2 \PQb jets.
The expected number of signal and background events in each category prior to the fit and the observed yields are given in Table~\ref{tab:yields_ljets}.
Good agreement is observed between data and expectations.

\begin{table}[!tb]
\centering
\topcaption{The number of expected background and signal events and the observed event yields in the different
  b tag categories for the $\Pe$+jets and $\mu$+jets analyses, prior to the fit.
  With the exception of the QCD multijet estimate,
  for which the total uncertainty is reported,
  the uncertainties reflect the statistical uncertainty in the simulated samples.
  \label{tab:yields_ljets}}
  \newcolumntype{x}{D{,}{\,\pm\,}{4.3}}
\resizebox{\textwidth}{!}{\begin{tabular}{lxx{c}@{\hspace*{5pt}}xx{c}@{\hspace*{5pt}}xx}
\multirow{3}{*}{Source} & \multicolumn{8}{c}{b tag category} \\
& \multicolumn{2}{c}{0 \PQb} && \multicolumn{2}{c}{1 \PQb} &&\multicolumn{2}{c}{$\geq$2 \PQb} \\\cline{2-3}\cline{5-6}\cline{8-9}
& \multicolumn{1}{c}{$\Pe$+jets} & \multicolumn{1}{c}{$\mu$+jets} && \multicolumn{1}{c}{$\Pe$+jets} & \multicolumn{1}{c}{ $\mu$+jets} && \multicolumn{1}{c}{$\Pe$+jets} & \multicolumn{1}{c}{$\mu$+jets} \\
\hline
$\PQt\PW$ & 3.03,0.02 & 5.6,0.03 && 2.49,0.02 & 4.5,0.03 && 0.39,0.01 & 0.67,0.01 \\
\PW+jets & 776,17 & 1704,26 && 13,2 & 26,3 && 0.2,0.3 & 0.8,0.6 \\
Z/$\gamma^{*}$ & 136,4 & 162,5 && 1.7,0.5 & 2.8,0.6 && 0.1,0.1 & 0.1,0.1 \\
WV & 0.52,0.01 & 1.01,0.02 && \multicolumn{1}{c}{$<$0.01} & \multicolumn{1}{c}{$<$0.02} && \multicolumn{1}{c}{$<$0.01} & \multicolumn{1}{c}{$<$0.01} \\
QCD multijet & 440,130 & 490,150 && 3.6,1.1 & 28,8 && 2.5,0.8 & 2.0,0.8 \\[\cmsTabSkip]
\ttbar signal & 22.8,0.3 & 42.3,0.4 && 36.9,0.4 & 71.1,0.5 && 13.8,0.2 & 27.0,0.3 \\[\cmsTabSkip]
Total & 1380,130 & 2410,150 && 57.7,2.4 & 131,9 && 16.8,0.9 & 31,1 \\[\cmsTabSkip]
Observed data & \multicolumn{1}{c}{1375} & \multicolumn{1}{c}{2406} && \multicolumn{1}{c}{61} & \multicolumn{1}{c}{129} && \multicolumn{1}{c}{19} & \multicolumn{1}{c}{33} \\
\end{tabular}}
\end{table}

The $M(j,j')$ and \minDeltaR distributions are shown in Fig.~\ref{fig:ljetsdata}.
The distributions have been combined for the $\Pe$+jets and $\mu$+jets channels to maximize the statistical precision
and are shown for events  with different \PQb-tagged jet multiplicities.
Fair agreement is observed between data and the pre-fit expectations.

\begin{figure}[!tb]
\centering
\includegraphics[width=0.32\textwidth]{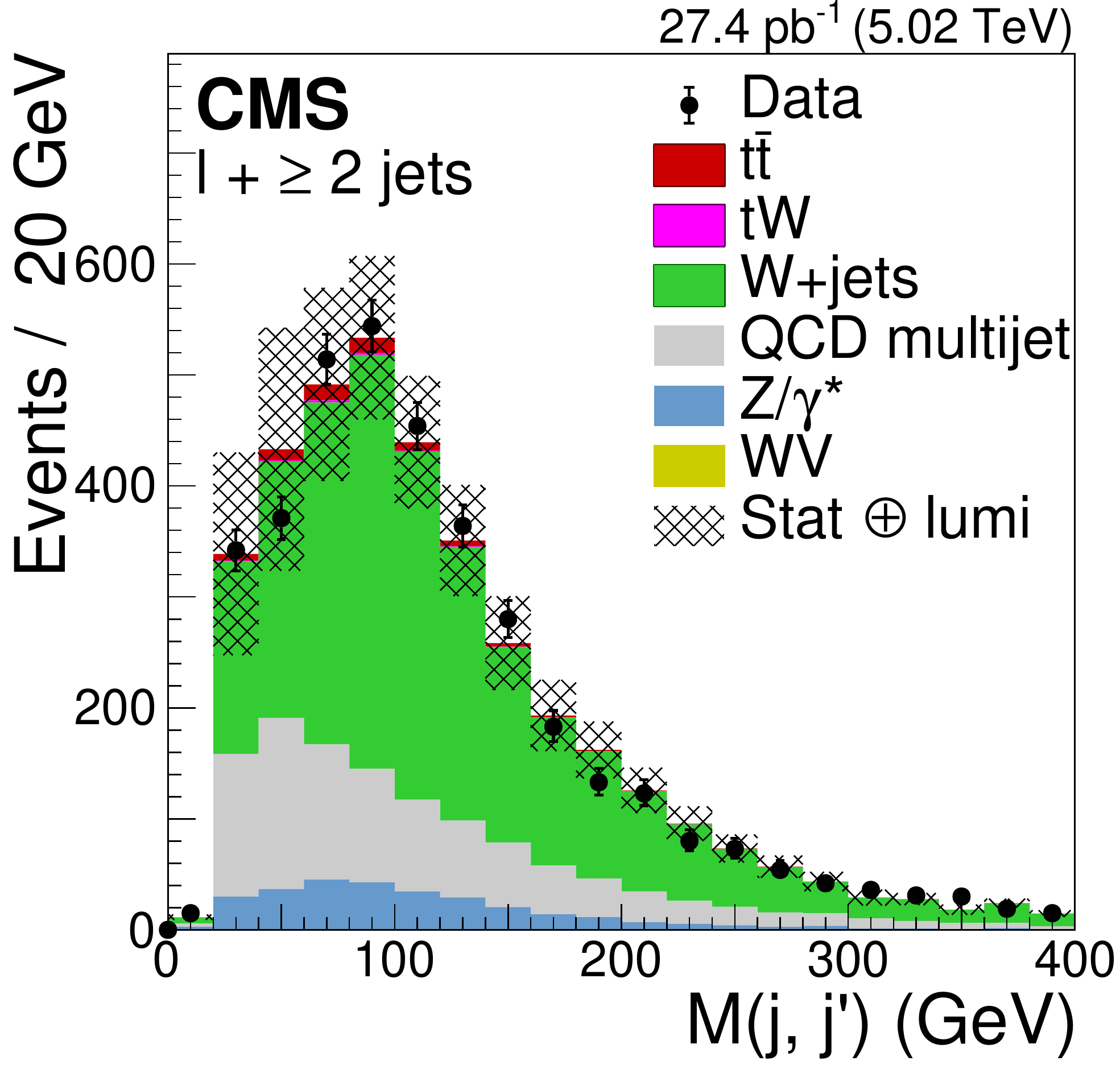}
\includegraphics[width=0.32\textwidth]{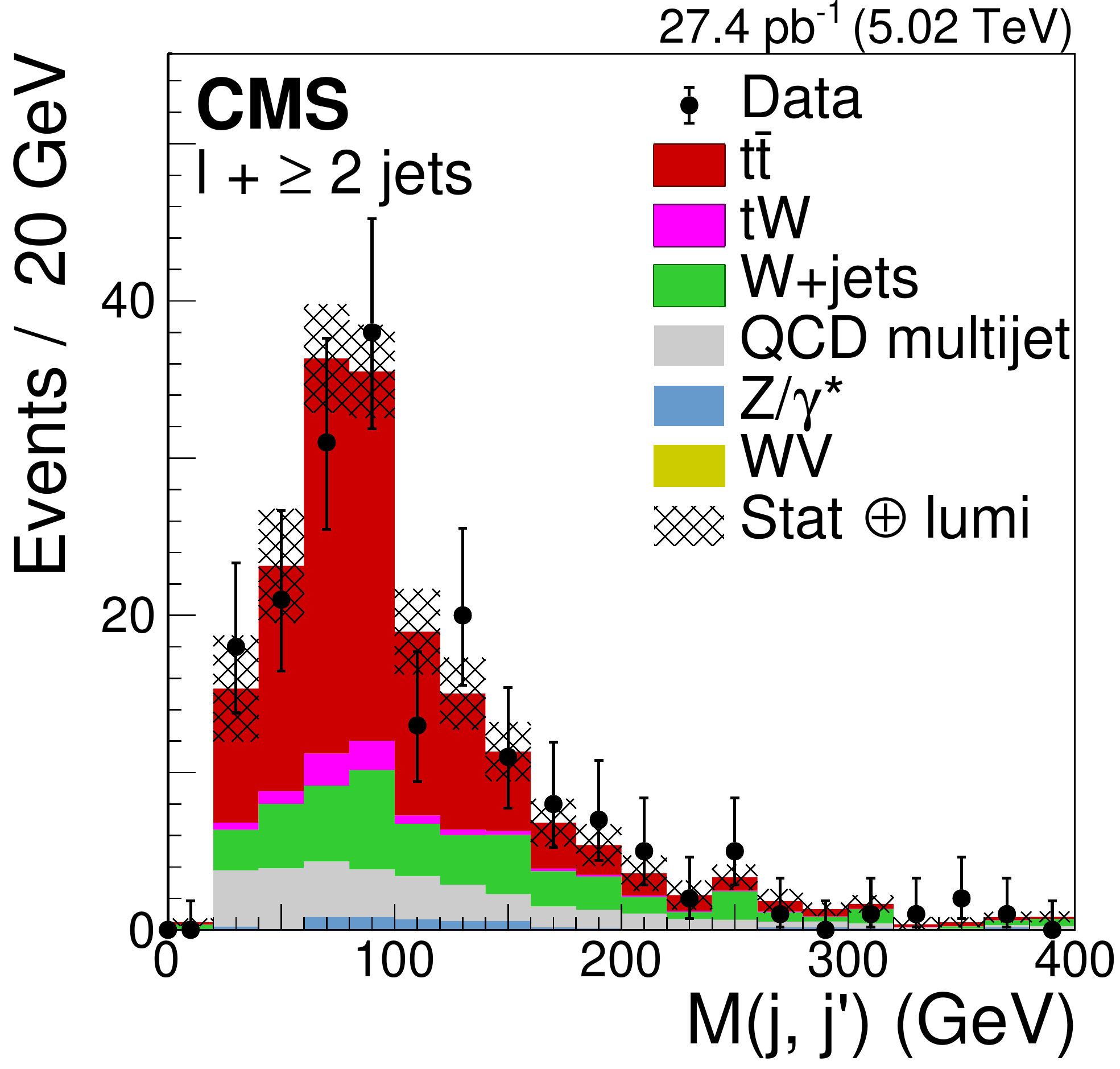}
\includegraphics[width=0.32\textwidth]{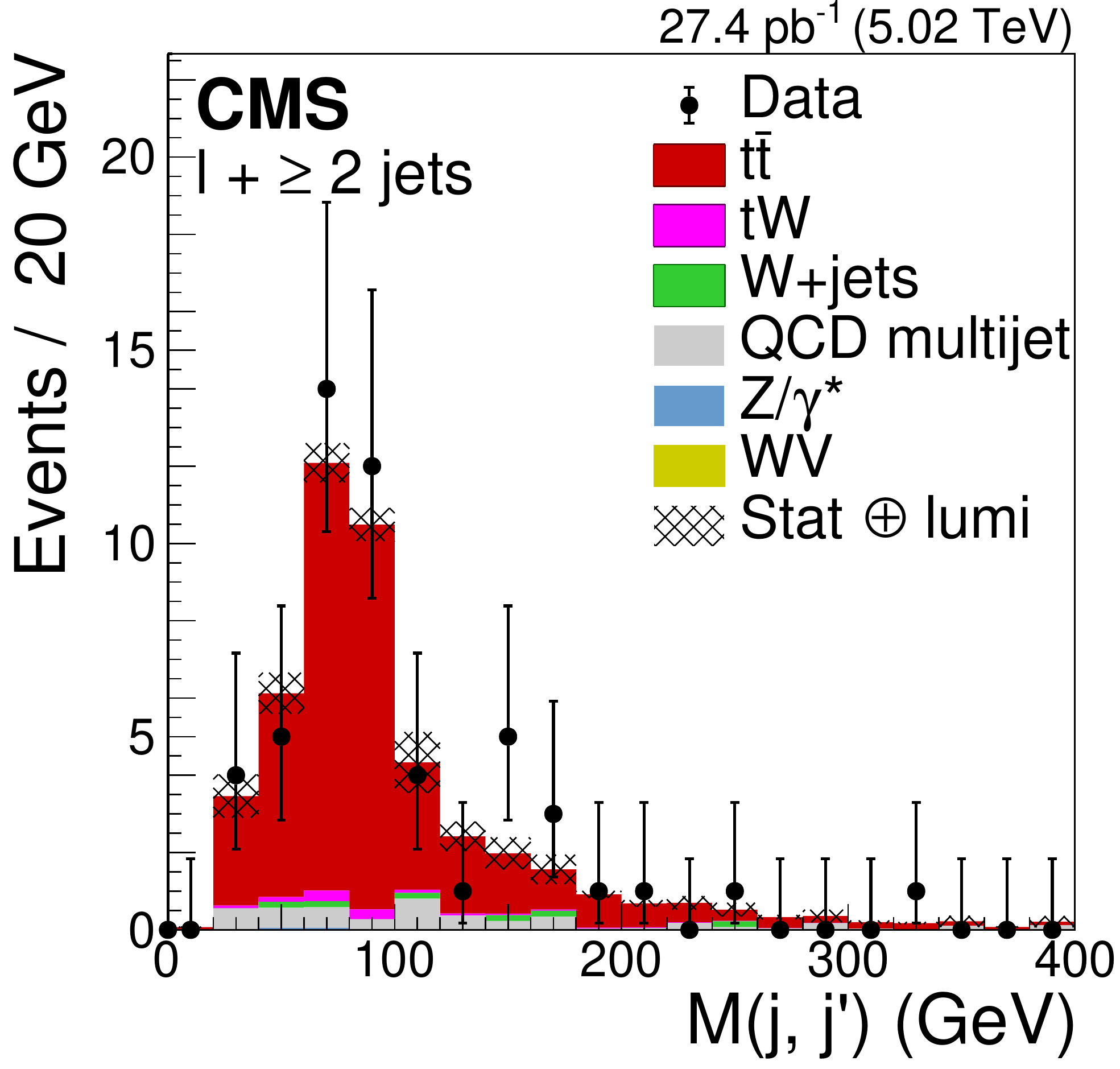}
\includegraphics[width=0.32\textwidth]{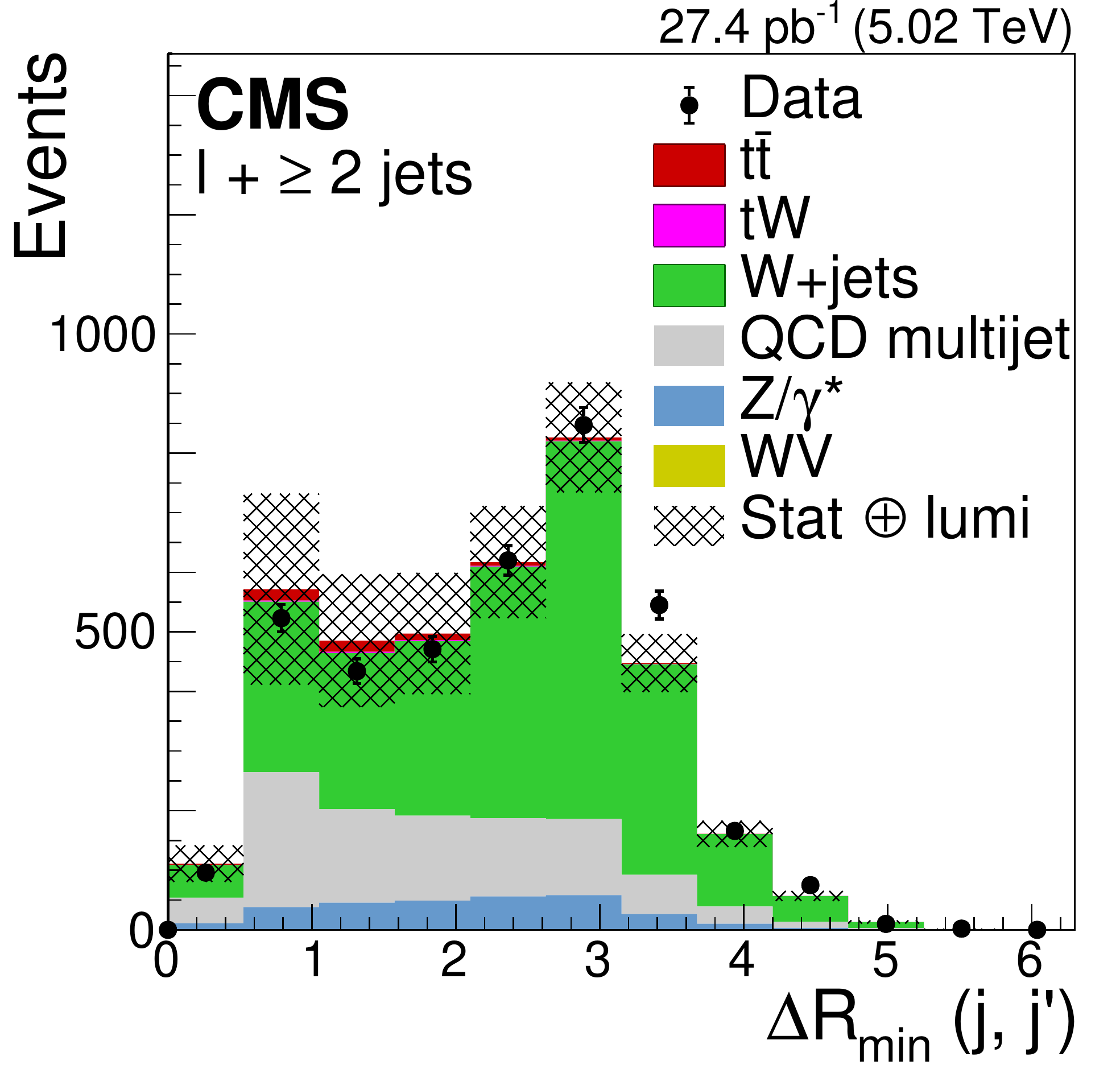}
\includegraphics[width=0.32\textwidth]{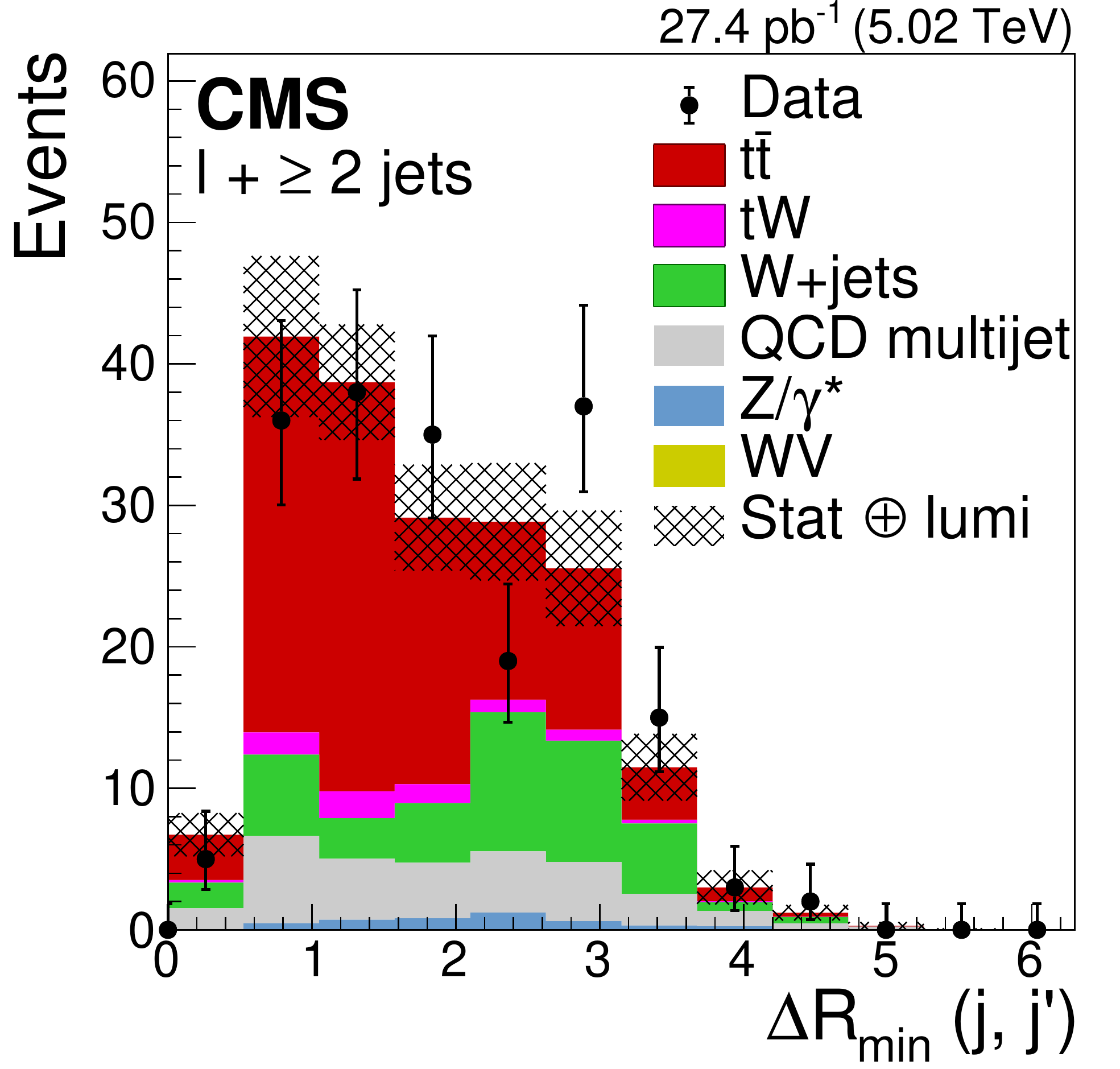}
\includegraphics[width=0.32\textwidth]{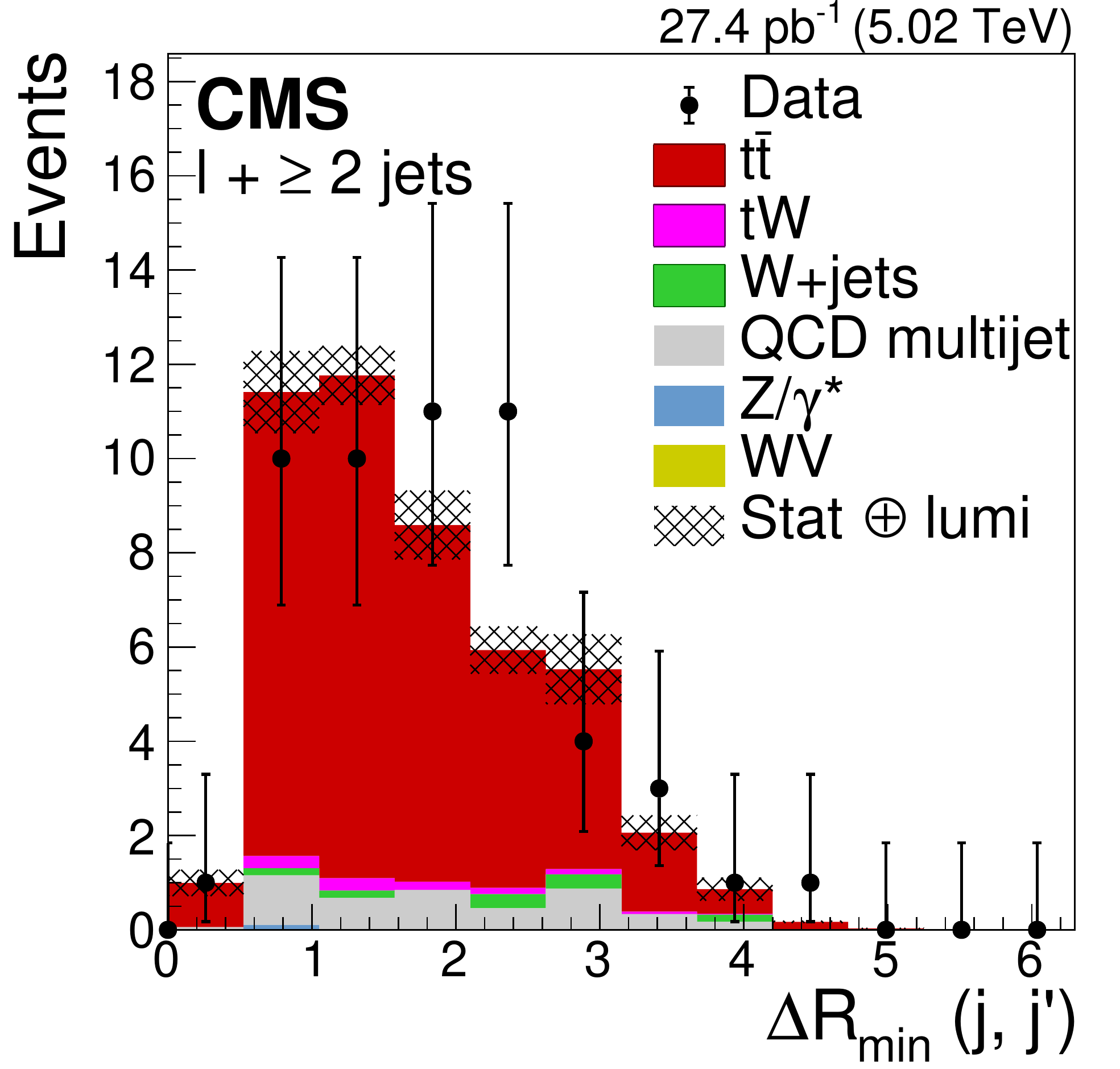}
\caption{
The predicted and observed distributions of the (upper row) $M(j,j')$ and (lower row) \minDeltaR variable for $\ell$+jets events
in the 0 \PQb (left), 1 \PQb (center), and $\geq$2 \PQb (right) tagged jet categories.
The distributions from data are compared to the sum of the expectations for the signal and backgrounds prior to any fit.
The QCD multijet background is estimated from data (see Section 5.1).
The cross-hatched band represents the statistical and the integrated luminosity uncertainties in the expected signal and background yields added in quadrature.
The vertical bars on the data points represent the statistical uncertainties.
}
\label{fig:ljetsdata}
\end{figure}

A profile likelihood ratio (PLR) method, similar to the one employed in Ref.~\cite{CMS-PAS-TOP-16-006}, is used to perform the fit.
In addition, a scale factor for the \PQb tagging efficiency ($\mathrm{SF}_{\PQb}$) is included as a parameter of interest in the fit.
The PLR is written as:
\begin{equation}
\lambda(\mu,\mathrm{SF}_{\PQb}) = \frac{\mathcal{L}(\mu,\mathrm{SF}_{\PQb},\hat{\hat\Theta})}{\mathcal{L}(\hat{\mu},\hat{\mathrm{SF}}_{\PQb},\hat\Theta)},
\label{eq:profll}
\end{equation}
where $\mu=\sigma/\sigma_\text{theo}$ is the signal strength (ratio of the observed \ttbar cross section to the expectation from theory)
and $\Theta$ is a set of nuisance parameters that encode the effect on the expectations due to variations in the sources of the systematic uncertainties
described in Section~\ref{sec:syst}.
The quantities $\hat{\hat\Theta}$ correspond to the values of the nuisance parameters that maximize the likelihood for the specified signal strength and \PQb tagging efficiency (conditional likelihood), and $\hat\mu$, $\hat{\mathrm{SF}}_{\PQb}$, $\hat\Theta$ are, respectively, the values of the signal strength, \PQb tagging efficiency, and nuisance parameters that maximize the likelihood.

Figure~\ref{fig:test} (left) shows the two-dimensional contours at the 68\% confidence level (CL)
obtained from the scan of $-2\ln(\lambda)$, as functions of $\mu$ and $\mathrm{SF}_{\PQb}$.
The expected results, obtained using the Asimov data set~\cite{Cowan:2010js},
are compared to the observed results and found to be in agreement well within one standard deviation.
The signal strength is obtained after profiling $\mathrm{SF}_{\PQb}$ and the result is
$\mu= 1.00\ ^{+0.10}_{-0.09} \stat\ ^{+0.09}_{-0.08}\syst$.
As a cross-check, the signal strength is also extracted by fitting only the total number of events observed in each of the six categories.
The observed value $\mu=1.03\ ^{+0.10}_{-0.10}\stat\ ^{+0.21}_{-0.11}\syst$ is in agreement with the analysis using the
\minDeltaR distributions.
Figure~\ref{fig:test} (right) summarizes the results obtained for the signal strength fit in each channel separately
from the analysis of the distributions and from event counting.
In both cases, a large contribution to the uncertainty is systematic in nature, although the statistical component is still significant.
In the  $\ell$+jets combination, the $\mu$+jets channel is expected and observed to carry the largest weight.

\begin{figure}[!tb]
\centering
\includegraphics[width=0.49\textwidth]{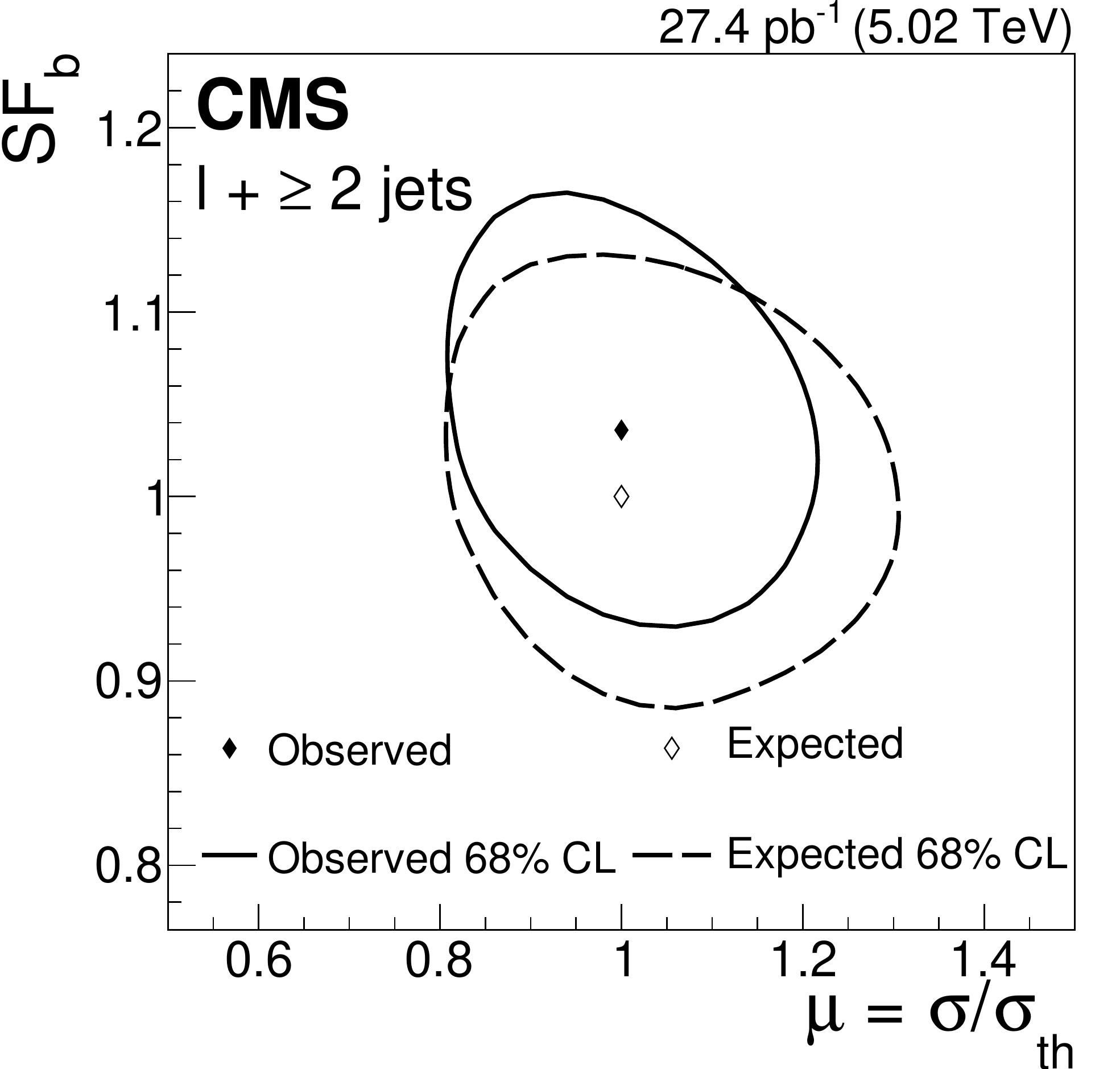}
\includegraphics[width=0.49\textwidth]{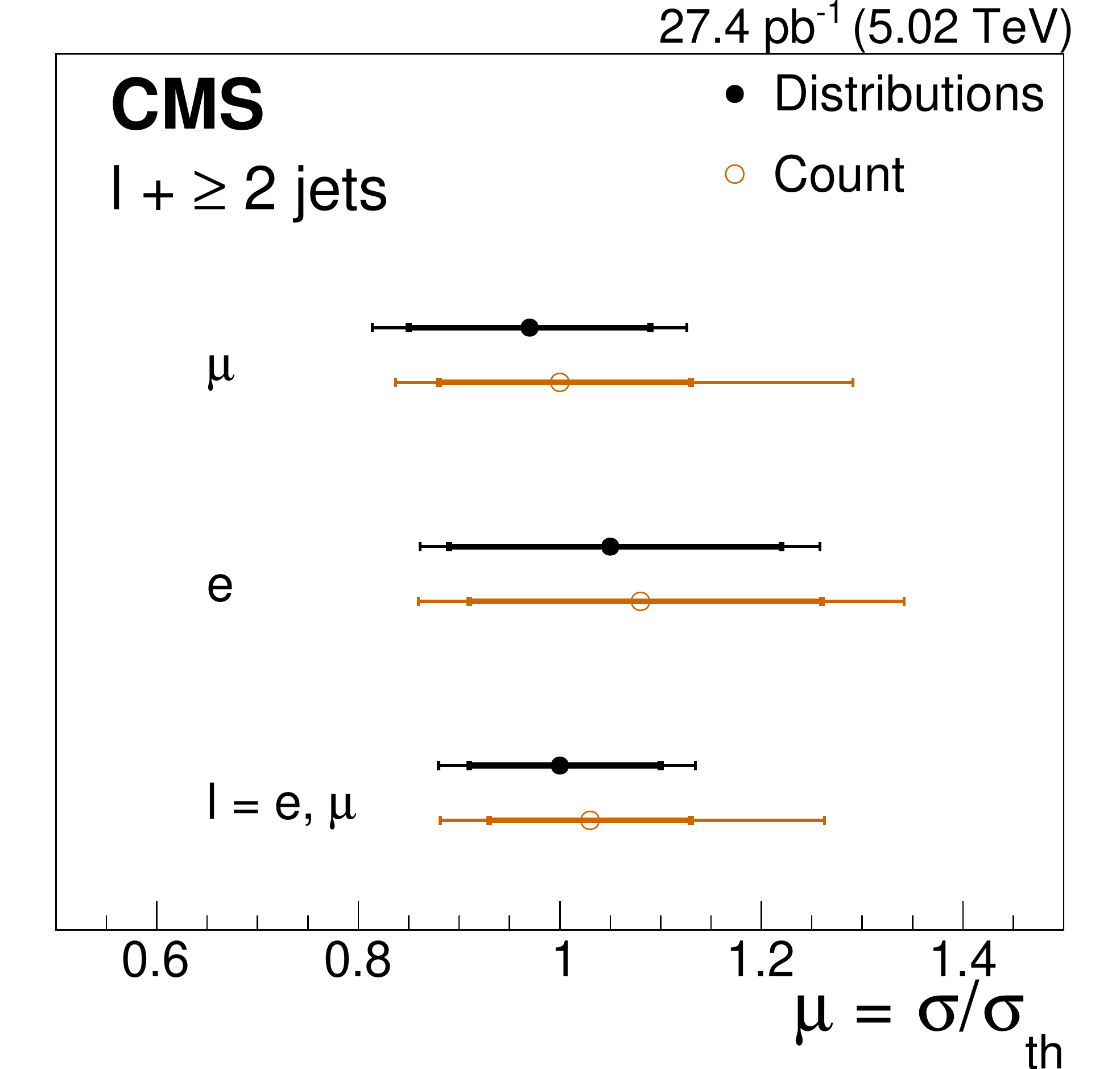}
\caption{Left: The 68\% CL contour obtained from the scan of the likelihood in $\ell$+jets analysis,
as a function of $\mu$ and $\mathrm{SF}_{\PQb}$ in the $\ell$+jets analysis.
The solid (dashed) contour refers to the result from data (expectation from simulation).
The solid (hollow) diamond represents the observed fit result (SM expectation).
Right: Summary of the signal strengths separately obtained in the $\Pe$+jets and $\mu$+jets channels, and after their combination in the $\ell$+jets channel.
The results of the analysis from the distributions are compared to those from the cross-check analysis with event counting (Count).
The inner (outer) bars correspond to the statistical (total) uncertainty in the signal strengths.
}
\label{fig:test}
\end{figure}

In order to estimate the impact of the experimental systematic uncertainties in the measured signal
strength, the fit is repeated after fixing one nuisance parameter at a time at its
post-fit uncertainty ($\pm$1 standard deviation) values. The impact on the signal strength fit is
then evaluated from the difference induced in the final result from
this procedure.
By repeating the fits, the effect of some nuisance parameters being fixed may be reabsorbed
by a variation of the ones being profiled, owing to correlations.
As such, the individual experimental uncertainties obtained and summarized
in Table~\ref{tab:impacts} can only be interpreted as the observed
post-fit values, and not as an absolute, orthogonalized breakdown of uncertainties.
With respect to the event counting, the analysis of the distributions
is less prone to the uncertainties in the QCD multijet background,
jet energy resolution, and signal modeling.
In both cases, the signal modeling uncertainties and the \PQb tagging efficiency are among the largest sources of uncertainty.

\begin{table}[!tb]
\centering
\topcaption{Estimated impact of each source of uncertainty in the
value of $\mu$ extracted from the analysis of distributions, and in the cross-check from event counting.
The ``Other background'' component includes the contributions from Z/$\gamma^{*}$, $\PQt\PW$, and WV events.
The total uncertainty is obtained by adding in quadrature the statistical, experimental systematic, and theoretical uncertainties.
The individual experimental uncertainties are obtained by repeating the fit after fixing one nuisance parameter at a time at its post-fit uncertainty ($\pm$1 standard deviation) value.
The values quoted have been symmetrized.\label{tab:impacts}}
\begin{tabular}{lll}
\multirow{2}{*}{Source} & \multicolumn{2}{c}{$\Delta\mu/\mu$} \\\cline{2-3}
                                            & Distr. & Count \\
\hline
Statistical uncertainty  & 0.095 & 0.100 \\
Experimental systematic uncertainty  & 0.085 & 0.160 \\[\cmsTabSkip]
\multicolumn{3}{c}{\textit{Individual experimental uncertainties}}\\
\hspace{+2mm} \PW{}+jets background  & 0.035 &  0.025 \\
\hspace{+2mm} QCD multijet background & 0.024 & 0.044 \\
\hspace{+2mm} Other background & 0.013 & 0.013 \\
\hspace{+2mm} Jet energy scale & 0.030 & 0.031 \\
\hspace{+2mm} Jet energy resolution & 0.006 & 0.023 \\
\hspace{+2mm} \PQb tagging & 0.034 & 0.045  \\
\hspace{+2mm} Electron efficiency & 0.011 & 0.028  \\
\hspace{+2mm} Muon efficiency & 0.017 & 0.022  \\[\cmsTabSkip]
\multicolumn{3}{c}{\textit{Theoretical uncertainties}}\\
Hadronization model of \ttbar signal & 0.028 & 0.069  \\
$\mu_\mathrm{R},\mu_\mathrm{F}$ scales of \ttbar signal (PS) & 0.044 & 0.115 \\
$\mu_\mathrm{R},\mu_\mathrm{F}$ scales of \ttbar signal (ME) & \hspace{-3.3mm}$<$0.010 & \hspace{-3.3mm}$<$0.010  \\[\cmsTabSkip]
Total uncertainty & 0.127 & 0.189 \\
\end{tabular}
\end{table}

The fiducial cross section is measured in events with one electron (muon) in the range $\pt>35$ (25)\GeV and $\abs{\eta}<2.1$ (including the transition region for electrons),
and at least two jets with $\pt>25\GeV$ and $\abs{\eta}<2.4$.
After multiplying the signal strength by the theoretical expectations (Eq.~(\ref{eq:theory})), we find
\begin{equation*}
\sigma_\text{fid}= 20.8 \pm 2.0\stat \pm 1.8\syst \pm 0.5\lum\unit{pb}.
\end{equation*}

The combined acceptance in the $\Pe$+jets and $\mu$+jets channels is estimated using the NLO \POWHEG simulation
to be  $\mathcal{A}=0.301\pm 0.007$, with the uncertainty being dominated by the variation of the $\mu_\mathrm{R},\mu_\mathrm{F}$ scales at ME and PS levels
and the hadronization model used for the \ttbar signal. The uncertainty due to the PDFs is included but verified to be less important.
Taking into account the acceptance of the analysis and its uncertainty, the inclusive \ttbar cross section is determined to be
\begin{equation*}
\stt= 68.9 \pm 6.5\stat \pm 6.1\syst \pm 1.6\lum\unit{pb},
\end{equation*}
in agreement with the SM prediction and attaining a 13\% total relative uncertainty.

\subsection{The dilepton final state}
\label{sec:res_dil}

In the dilepton analysis, the \ttbar cross section is extracted from an event counting measurement.
Figure~\ref{fig:jetsAnddilepton} shows the distributions of the jet multiplicity and the scalar \pt sum of all jets ($H_\mathrm{T}$),
for events passing the dilepton criteria in the \empm channel.
In addition, it displays the lepton-pair invariant mass and \pt distributions,
after requiring at least two jets in the event in the \empm channel.
Figure~\ref{fig:dimuon} shows the \ptmiss and the lepton-pair invariant mass distributions in the \mmpm channel for events passing
the dilepton criteria, and the Z boson veto with the $\ptmiss>35$\GeV requirement, in the second case.
The predicted distributions take into account the efficiency corrections described in
Section~\ref{sec:evtSel}
and the background estimations discussed in Section~\ref{sec:bkg_dil}.
Good agreement is observed between
the data and predictions for both signal and background.

\begin{figure}[!tb]
\centering
\includegraphics[width=0.49\textwidth]{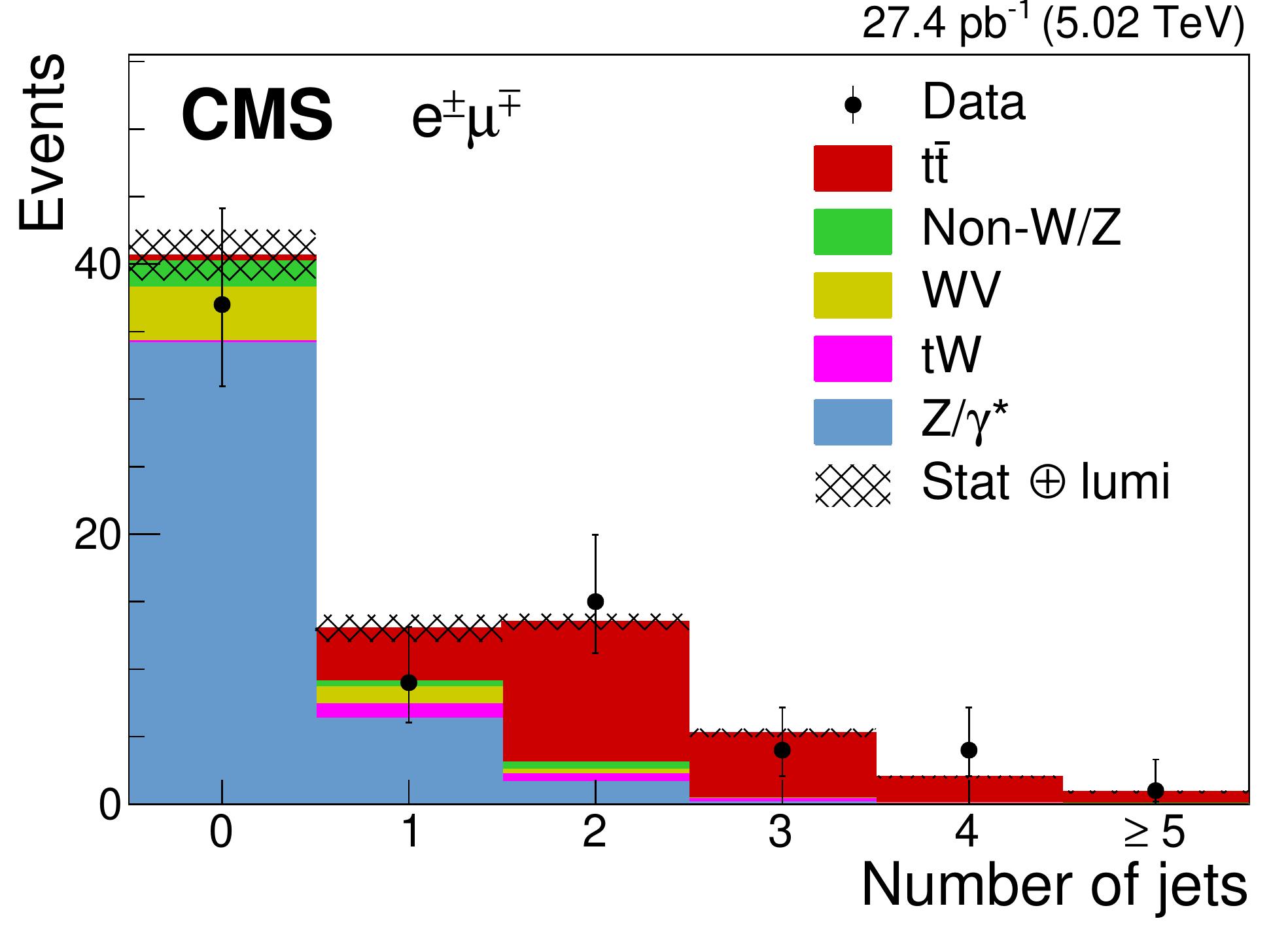}
\includegraphics[width=0.49\textwidth]{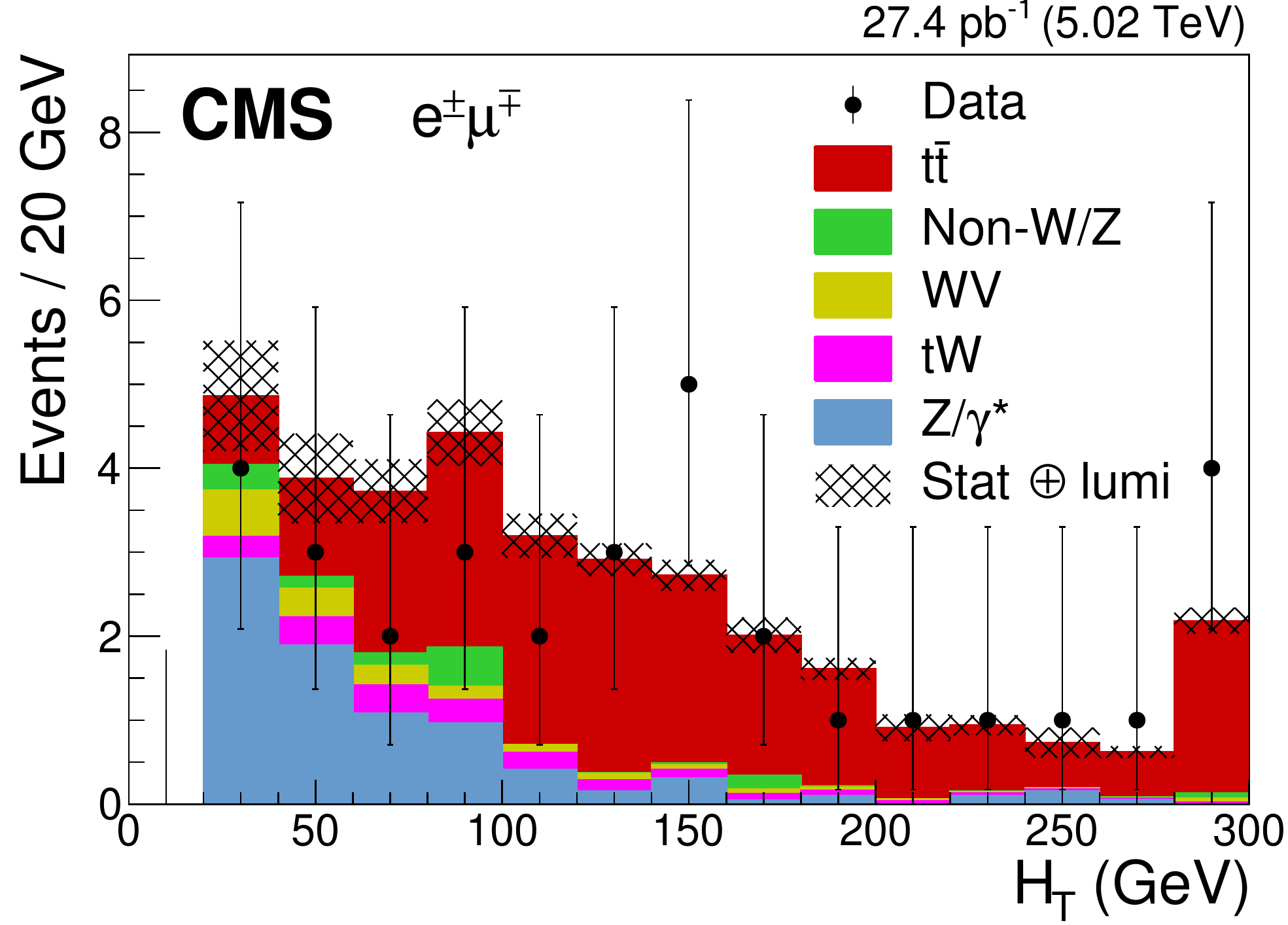}
\includegraphics[width=0.49\textwidth]{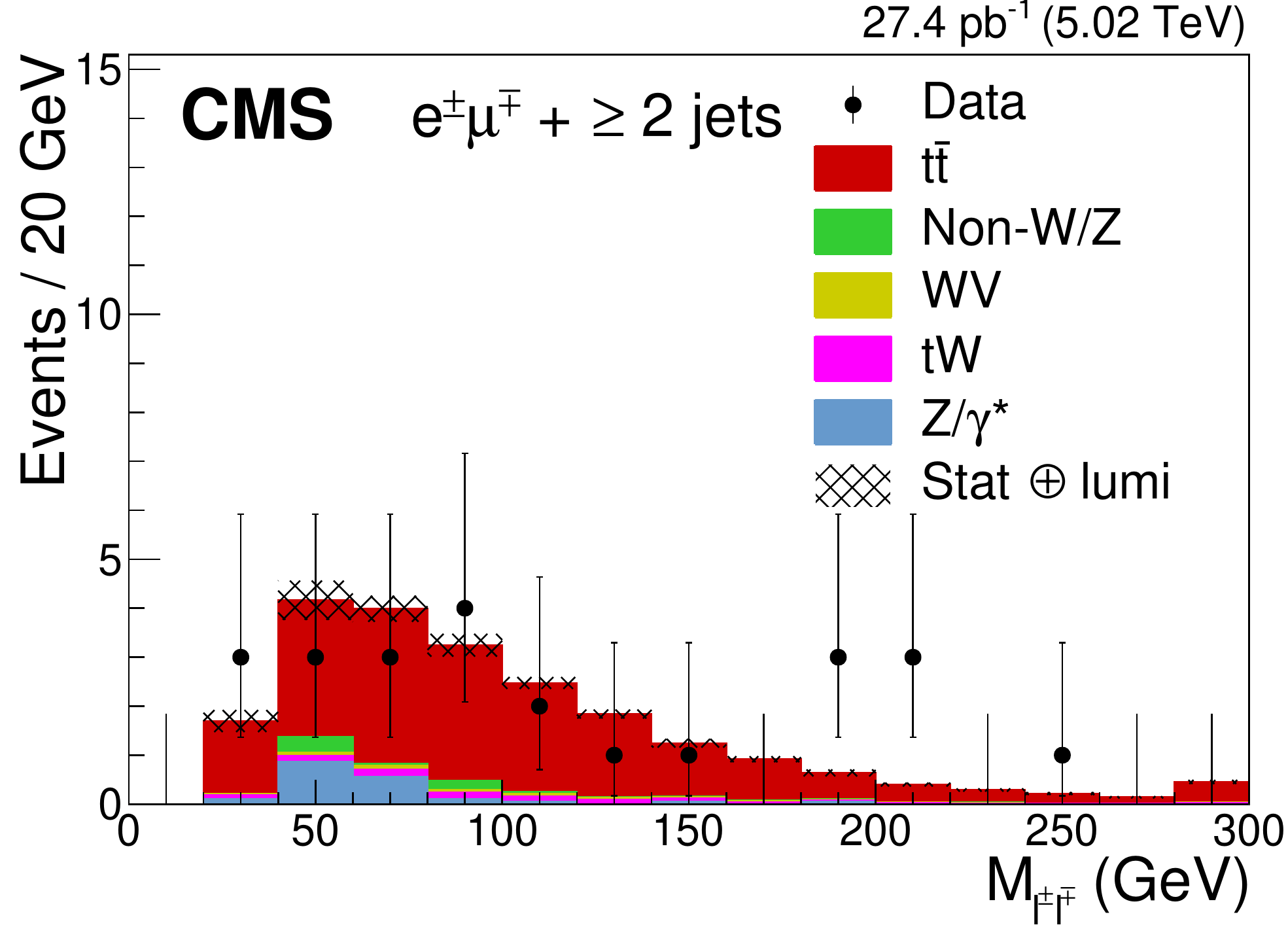}
\includegraphics[width=0.49\textwidth]{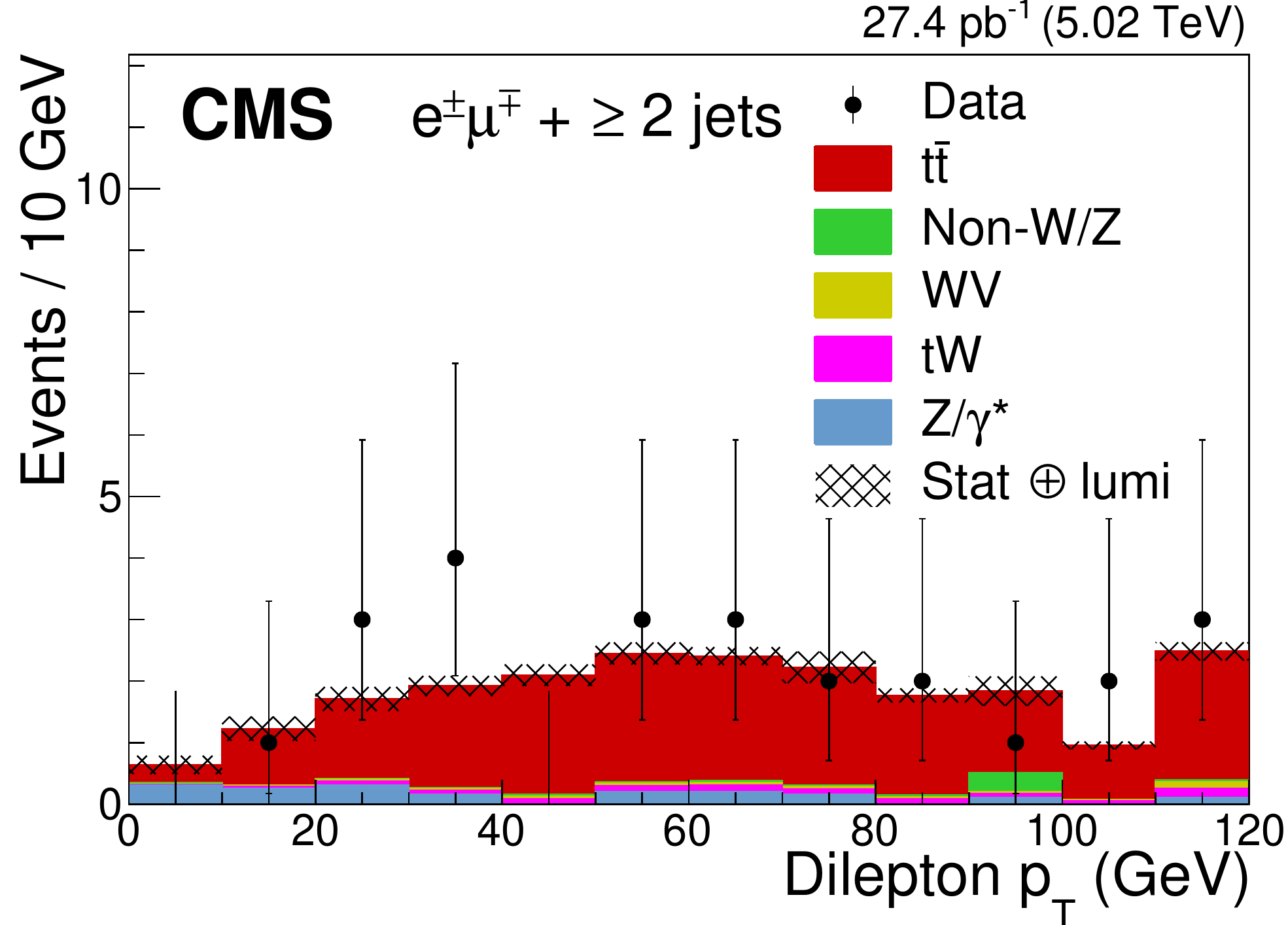}
\caption{
Predicted and observed distributions of the (upper row) jet multiplicity and scalar \pt sum of all jets ($H_\mathrm{T}$) for events passing the dilepton criteria, and of the (lower row)
invariant mass and \pt of the lepton pair after requiring at least two jets, in the \empm channel.
The Z/$\gamma^{*}$ and \nonW backgrounds are determined from data (see Section~\ref{sec:bkg_dil}).
The cross-hatched band represents the statistical and integrated luminosity uncertainties in the expected signal and background yields added in quadrature.
The vertical bars on the data points represent the statistical uncertainties.
The last bin of the distributions contains the overflow events.
}
\label{fig:jetsAnddilepton}
\end{figure}

\begin{figure}[!tb]
\centering
\includegraphics[width=0.49\textwidth]{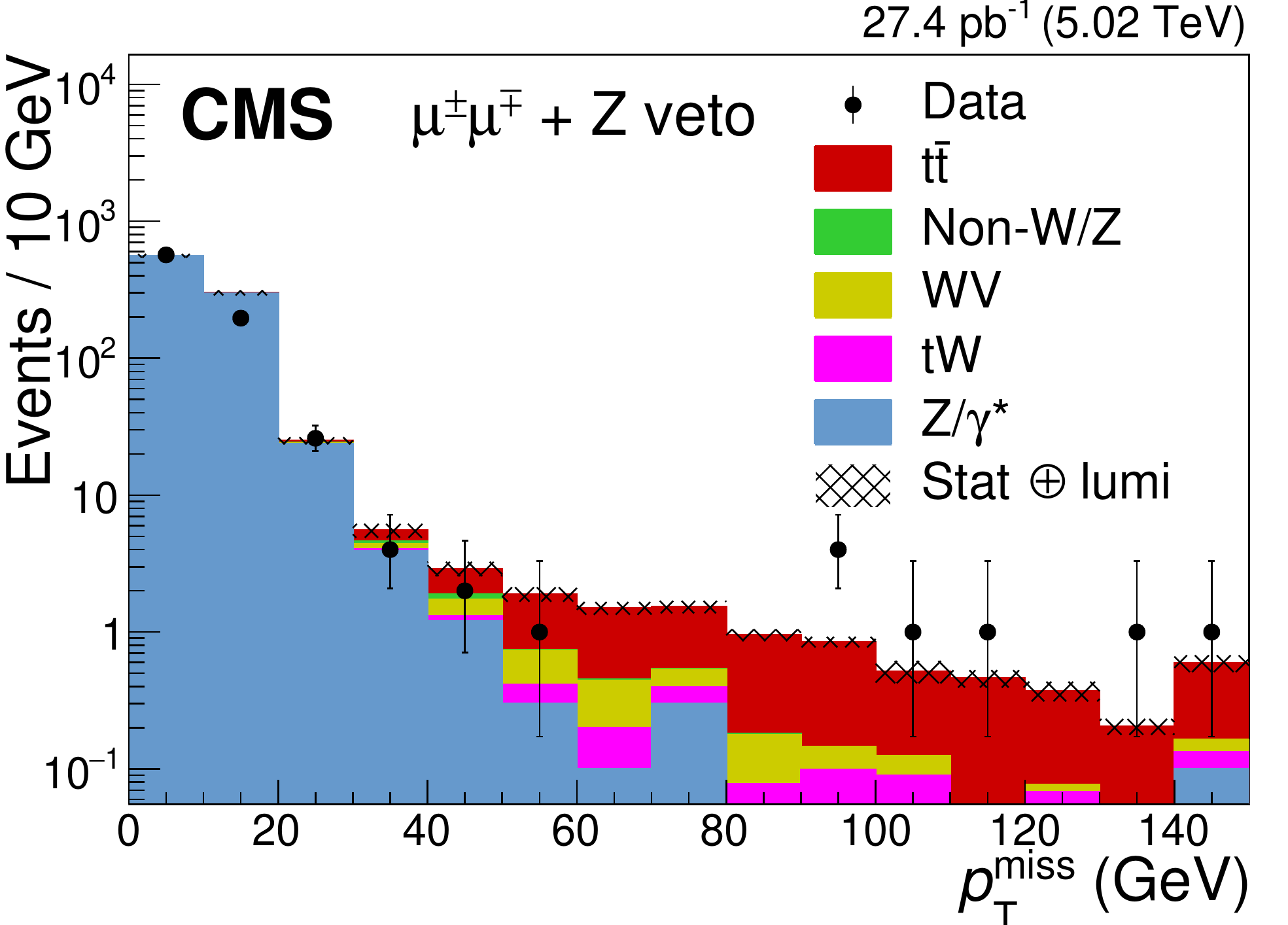}
\includegraphics[width=0.49\textwidth]{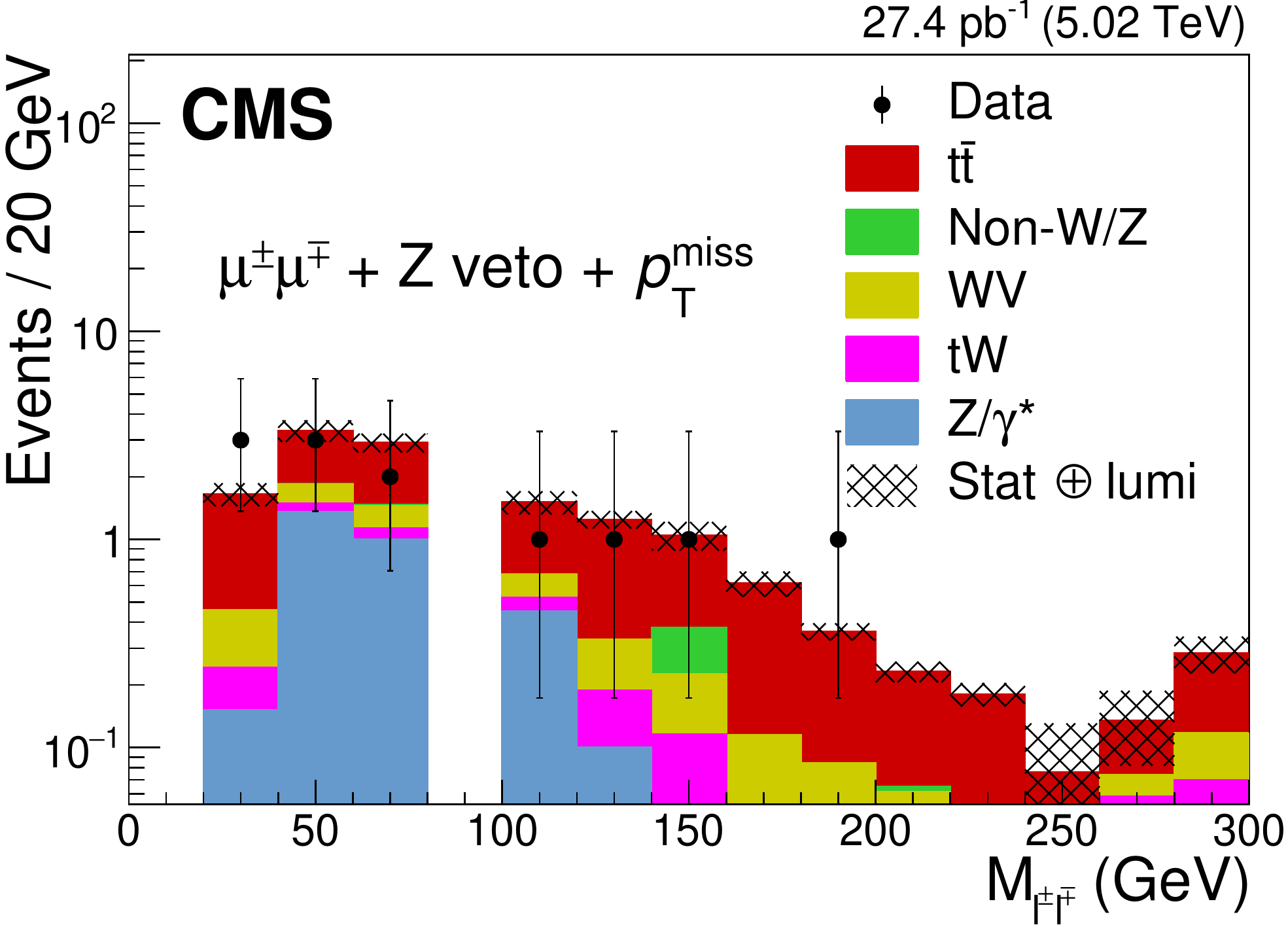}
\caption{
Predicted and observed distributions of the (left) \ptmiss in events passing the dilepton criteria and Z boson veto, and of the (right) invariant mass
of the lepton pair after the $\ptmiss>35$\GeV requirement in the \mmpm channel.
The cross-hatched band represents the statistical and integrated luminosity uncertainties in the expected signal and background yields added in quadrature.
The vertical bars on the data points represent the statistical uncertainties.
The last bin of the distributions contains the overflow events.
}
\label{fig:dimuon}
\end{figure}

The fiducial \ttbar production cross section is measured by counting events in the visible phase space (defined by the same $\pt$, $\abs{\eta}$, and multiplicity requirements
for leptons and jets as described in Section~\ref{sec:evtSel}, but including the transition region for electrons) and is denoted by $\sigma_\text{fid}$.
It is extrapolated to the full phase space in order to determine the inclusive \ttbar cross section using the expression
\begin{equation}
\stt = \frac{N - N_\mathrm{B}}{\varepsilon \, \mathcal{A} \, {\mathcal{L}}} = \frac{\sigma_\text{fid}}{\mathcal{A}},
\label{eqn:xsecA}
\end{equation}
where $N$ is the  total number of dilepton events observed in data, $N_\mathrm{B}$ the
number of estimated background events,
$\varepsilon$ the selection efficiency,
$\mathcal{A}$ the acceptance,
 and $\mathcal{L}$ the integrated luminosity. Table~\ref{tab:yields_dilepton} gives the total number of events observed in
data, together with the total number of signal and background events expected
from simulation or estimated from data, after the full set of selection criteria.
The total detector, trigger, and reconstruction efficiency is estimated from data to be $\varepsilon = 0.55\pm0.02$ ($0.57\pm0.04$) in the \empm (\mmpm) channel.
Using the definitions above, the yields from Table~\ref{tab:yields_dilepton}, and the systematic uncertainties from Table~\ref{tab:breakdown_comb}, the measured fiducial cross section for \ttbar production is
\begin{equation*}
\sigma_\text{fid}  =  41 \pm 10\stat \pm 2\syst \pm1\lum\unit{pb}
\end{equation*}
in the \empm\ channel and
\begin{equation*}
\sigma_\text{fid}  =  22\pm 11\stat \pm 4\syst \pm1\lum\unit{pb}
\end{equation*}
in the \mmpm\ channel.

\begin{table}[!tb]
\centering
\topcaption{The predicted and observed numbers of dilepton events obtained after applying the full selection.
The values are given for the individual sources of background, \ttbar signal, and data.
The uncertainties correspond to the statistical component.
}
\begin{tabular}{lcc}
Source                  & \empm\ & \mmpm      \\
\hline
  $\PQt\PW$            & 0.92 $\pm$ 0.02  &   0.29 $\pm$ 0.01\\
   Non-W/Z leptons      & 1.0  $\pm$ 0.9   &   0.04 $\pm$ 0.01\\
   Z/$\gamma^{*}$       & 1.6  $\pm$ 0.2   &   1.1  $\pm$ 0.8\\
  \WV                   & 0.44 $\pm$ 0.02  &   0.15 $\pm$ 0.01\\[\cmsTabSkip]
  \ttbar signal         & 18.0 $\pm$ 0.3   &   6.4 $\pm$ 0.2  \\[\cmsTabSkip]
  Total                 & 22.0  $\pm$ 0.9   &  7.9 $\pm$ 0.8\\[\cmsTabSkip]
  Observed data         & 24		   &   7\\
\end{tabular}
\label{tab:yields_dilepton}
\end{table}

\begin{table}[!tb]
\topcaption{Summary of the individual contributions to the systematic
uncertainty in the $\stt$ measurements for the dilepton channels. The relative uncertainties $\Delta\stt /\stt$ (in \%),
as well as absolute uncertainties in \stt, $\Delta\stt$ (in pb), are presented. The statistical and total uncertainties are also given,
where the latter are the quadrature sum of the statistical and systematic uncertainties.}
\centering
\begin{tabular}{lcc{c}@{\hspace*{5pt}}cc}

                                     & \multicolumn{2}{c}{\empm}  && \multicolumn{2}{c}{\mmpm}   \\\cline{2-3}\cline{5-6}
Source                               & $\Delta\stt /\stt$ (\%)  & $\Delta\stt$ (pb)  &&  $\Delta\stt /\stt$ (\%)  &  $\Delta\stt$ (pb) \\
\hline
Electron efficiency                & 1.4  & 1.0   &&---  &  --- \\
Muon efficiency                    & 3.0  & 2.3   && 6.1  &  3.6 \\
Jet energy scale                     & 1.3  & 1.0   && 1.3  &  0.7 \\
Jet energy resolution                & \hspace{-2.8mm}$<$0.1 & \hspace{-2.8mm}$<$0.1  && \hspace{-2.8mm}$<$0.1 & \hspace{-2.8mm}$<$0.1 \\
Missing transverse momentum          & ---  & ---   &&0.7  &  0.4 \\
$\mu_\mathrm{R},\mu_\mathrm{F}$ scales of \ttbar signal (PS)     & 1.2  & 0.9   && 1.7  & 1.0 \\
$\mu_\mathrm{R},\mu_\mathrm{F}$ scales of \ttbar signal (ME)     & 0.2  & 0.1   && 1.1  & 0.6 \\
Hadronization model of \ttbar signal & 1.2  & 0.9   && 5.2  & 3.1 \\
PDF                                  & 0.5  & 0.4   && 0.4  & 0.2 \\
MC sample size & 1.4  & 1.1   && 2.4  & 1.4 \\
$\PQt\PW$ background                 &  1.4 & 1.1   && 1.6  & 0.9 \\
WV background                        &  0.7 & 0.5   && 0.9  & 0.5 \\
Z/$\gamma^{*}$ background            &  2.7 & 2.1   && 15   & 9.1 \\
Non-W/Z background                   &  2.5 & 1.9   && 0.7  & 0.4 \\[\cmsTabSkip]
Total systematic uncertainty         & \multirow{2}{*}{ 5.8 } & \multirow{2}{*}{ 4.4 }  && \multirow{2}{*}{ 18 }  & \multirow{2}{*}{ 11 } \\
(w/o integrated luminosity)          &        &        &&      &  \\[\cmsTabSkip]
Integrated luminosity                &   2.3  &   1.8  &&  2.3 &  1.4 \\[\cmsTabSkip]
Statistical uncertainty              &  25  &  19  && 48 & 29 \\[\cmsTabSkip]
Total uncertainty                    &  25  &  19  && 52  & 31 \\
\end{tabular}
\label{tab:breakdown_comb}
\end{table}

The acceptance, as estimated from MC simulation, is found to be $\mathcal{A} = 0.53\pm0.01$ ($0.37\pm0.01$) in the \empm (\mmpm) channel. The statistical uncertainty (from MC simulation) is included in the uncertainty in $\mathcal{A}$. By extrapolating to the full phase space, the inclusive \ttbar cross section is measured to be
\begin{equation*}
\stt  =  77 \pm 19\stat \pm  4\syst \pm 2\lum\unit{pb}
\end{equation*}
in the \empm channel and
\begin{equation*}
\stt  =  59 \pm 29\stat \pm 11\syst \pm 1\lum\unit{pb}
\end{equation*}
in the \mmpm\ channel.
Table~\ref{tab:breakdown_comb} summarizes the
relative and absolute statistical and
systematic uncertainties from different sources contributing to \stt.
The separate total systematic uncertainty without the uncertainty in the integrated luminosity, the part attributed to the integrated luminosity, and the statistical
contribution are added in quadrature to obtain the total uncertainty.
The cross sections, measured with a relative uncertainty of 25 and 52\%, are in agreement with the SM
prediction (Eq.~(\ref{eq:theory})) within the uncertainties in the measurements.
\subsection{Combination}
\label{sec:comb}

The three individual \stt measurements are combined using the BLUE method~\cite{Lyons:1988rp,Valassi:2013bga} to determine an overall \ttbar cross section.
All systematic uncertainties are considered as fully correlated across all channels, with the following exceptions:
the uncertainty associated with the finite event size of the simulated samples is taken as uncorrelated;
the electron identification is not relevant for the $\mu\mu$ channel;
and the \PQb{} tagging and QCD multijet background uncertainties are only considered for the $\ell$+jets channel.
In the $\ell+$jets channel, the WV and $\cPZ/\gamma^*$ backgrounds are not considered separately but as part of the ``Other backgrounds'' component, which is dominated by tW events.
The uncertainty associated with this category is therefore treated as fully correlated with the tW uncertainty in the dileptonic channels and uncorrelated with the WV and $\cPZ/\gamma^*$ uncertainties.

The combined inclusive \ttbar cross section is measured to be:
\begin{equation*}
 \stt  =  69.5 \pm 6.1\stat \pm 5.6\syst \pm 1.6\lum\unit{pb}=69.5\pm8.4\,\text{(total)}\unit{pb},
\end{equation*}
where the total uncertainty is the sum in quadrature of the individual uncertainties. The weights of the individual measurements, to be understood in the sense of Ref.~\cite{Valassi:2013bga}, are 81.8\% for $\ell+$jets, 13.5\% for \empm, and 4.7\% for \mmpm channels.

The combined result is found to be robust by performing an iterative variant of the BLUE method~\cite{Lista:2014qia} and varying some assumptions on the correlations of different combinations of systematic uncertainties. Also, the post-fit correlations between the nuisance parameters in the $\ell$+jets channel have been checked and found to have negligible impact.

Figure~\ref{fig:sqrt} presents a summary of CMS measurements~\cite{Khachatryan:2016mqs,Khachatryan:2016yzq,CMS-PAS-TOP-16-005,CMS-PAS-TOP-16-006}
of $\sigma_{\ttbar}$ in $\Pp\Pp$ collisions at different \sqrts in the $\ell$+jets and dilepton channels, compared to the NNLO+NNLL
prediction using the NNPDF3.0 PDF set with $\alpha_\mathrm{s}(M_{\Z})=0.118$ and $m_\text{top} = 172.5\GeV$.
In the inset, the results from this analysis at $\sqrts = 5.02$\TeV are also compared to the predictions from the MMHT14~\cite{Harland-Lang:2014zoa}, CT14~\cite{Dulat:2015mca}, and
ABMP16~\cite{Alekhin:2017kpj}
PDF sets, with the latter using $\alpha_\mathrm{s}(M_{\Z})=0.115$ and $m_\text{top} = 170.4\GeV$.
Theoretical predictions using different PDF sets have comparable values and uncertainties, once consistent values of $\alpha_\mathrm{s}$ and $m_\text{top}$ are associated with the respective PDF set.

\begin{figure*}[!tb]
\centering
{\includegraphics[width=0.98\textwidth]{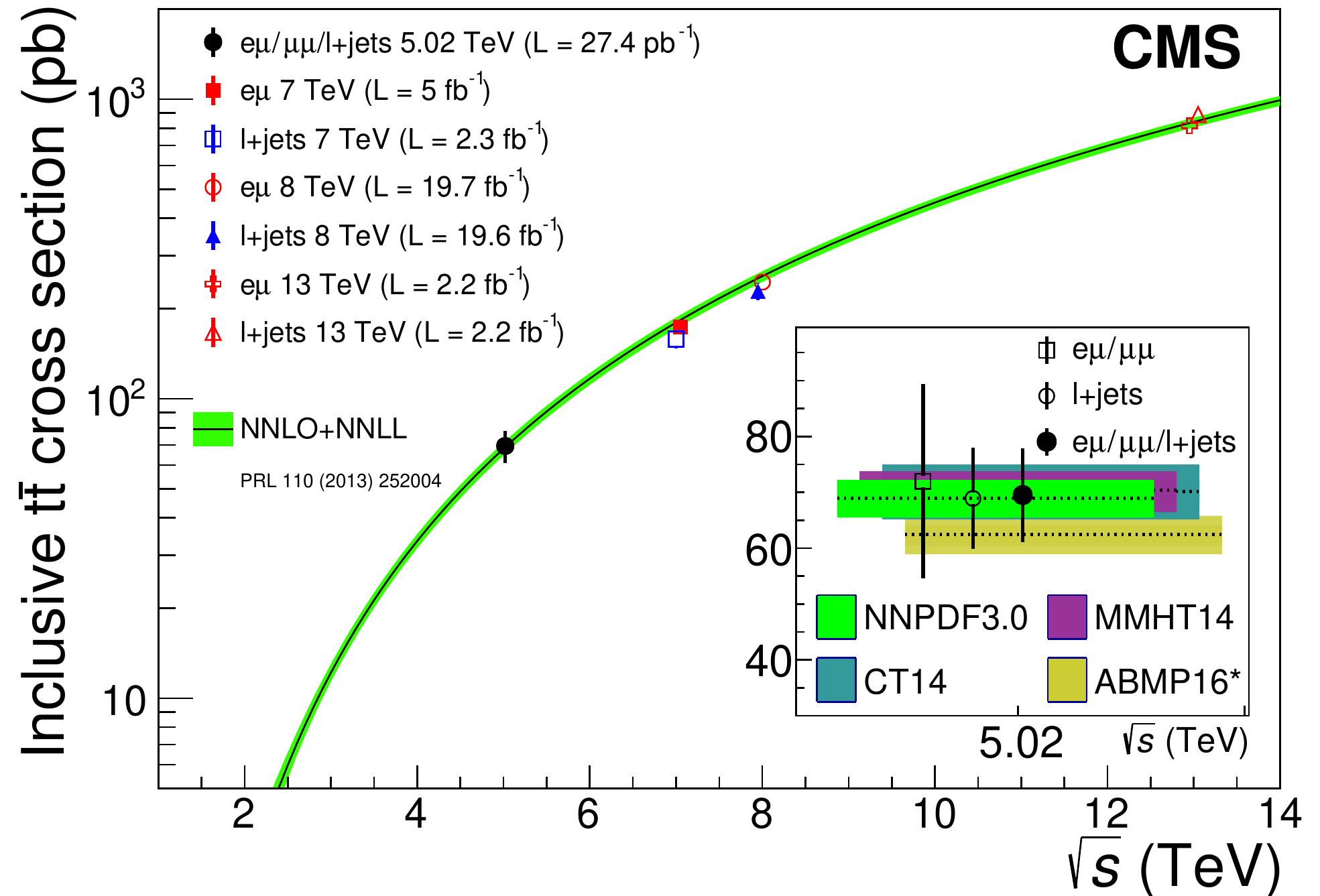}}
\caption{
  Inclusive \stt in
  pp collisions as a
  function of the center-of-mass energy;
  previous CMS measurements at $\sqrts = 7$, 8~\cite{Khachatryan:2016mqs,Khachatryan:2016yzq}, and 13~\cite{CMS-PAS-TOP-16-005,CMS-PAS-TOP-16-006}\TeV in the separate $\ell$+jets and dilepton channels are displayed, along with the combined measurement at 5.02\TeV from this analysis.
  The NNLO+NNLL theoretical prediction~\cite{mitov} using the NNPDF3.0~\cite{Ball:2014uwa} PDF set with $\alpha_\mathrm{s}(M_{\Z})=0.118$ and $m_\text{top} = 172.5\GeV$ is shown in the main plot.
  In the inset, additional predictions at $\sqrts = 5.02$\TeV using the MMHT14~\cite{Harland-Lang:2014zoa}, CT14~\cite{Dulat:2015mca}, and ABMP16~\cite{Alekhin:2017kpj} PDF sets,
  the latter with $\alpha_\mathrm{s}(M_{\Z})=0.115$ and $m_\text{top} = 170.4\GeV$, are compared, along with the NNPDF3.0 prediction, to the individual and combined results from this analysis.
  The vertical bars and bands represent the total uncertainties in the data and in the predictions, respectively.
}
\label{fig:sqrt}
\end{figure*}
\section{QCD analysis}
\label{sec:pdfs}

To illustrate the impact of the \stt measurements at $\sqrts = 5.02$\TeV on the knowledge of the proton PDFs, the results are used in a QCD analysis at
NNLO, together with the combined measurements of neutral- and charged-current cross sections for deep inelastic electron- and positron-proton
scattering (DIS) at HERA~\cite{Abramowicz:2015mha}, and the CMS measurement~\cite{Khachatryan:2016pev} of the muon charge asymmetry in $\PW$ boson
production at $\sqrt{s}=8$\TeV. The latter data set is used in order to improve the constraint on the light-quark distributions.

Version 2.0.0 of \textsc{xFitter}~\cite{Alekhin:2014irh, herafitter}, the open-source QCD-analysis framework for PDF
determination, is employed, with the partons evolved using the Dokshitzer--Gribov--Lipatov--Altarelli--Parisi
equations~\cite{Gribov:1972ri,Altarelli:1977zs,Curci:1980uw,Furmanski:1980cm,Moch:2004pa,Vogt:2004mw}
at NNLO, as implemented in the \textsc{qcdnum} 17-01/13 program~\cite{Botje:2010ay}. The treatment and the choices for the central values
and variations of the c and b quark masses, the strong coupling, and the strange-quark content fraction of the proton follow that of
earlier CMS analyses, e.g., Ref.~\cite{Khachatryan:2016pev}. The $\mu_\mathrm{R},\mu_\mathrm{F}$ scales are set to the four-momentum transfer
in the case of the DIS data, the $\PW$ boson mass for the muon charge asymmetry results, and the top quark mass in the case of \stt.

The systematic uncertainties in all three measurements of \stt and their correlations are treated the same way as in
the combination described in Section~\ref{sec:comb}. The theoretical predictions for \stt are obtained at NNLO using the
\textsc{hathor} calculation~\cite{Aliev:2010zk}, assuming $m_\text{top}= 172.5\GeV$.
The bin-to-bin correlations of the experimental uncertainties in the muon charge asymmetry and DIS
measurements are taken into account. The theoretical predictions for the muon charge asymmetry are obtained as
described in Ref.~\cite{Khachatryan:2016pev}.

The procedure for the determination of the PDFs follows the approach used in the QCD ana\-ly\-sis of Ref.~\cite{Khachatryan:2016pev} and
results in a 14-parameter fit. The parametrized PDFs are the gluon distribution, $x\Pg$, the valence quark distributions, $x\PQu_v$, $x\PQd_v$, and
the $\PQu$-type and $\PQd$-type antiquark distributions, $x\overline{U}$, $x\overline{D}$. The relations $x\overline{U} = x\cPaqu$
and $x\overline{D} = x\PAQd + x\PAQs$ are assumed at the initial scale of the QCD evolution $Q_0^2 = 1.9\GeV^2$. At this scale, the parametrizations are of the form:
\begin{align}
x\Pg(x) &= A_\Pg x^{B_\Pg}\,(1-x)^{C_{\Pg}}\, (1+D_\Pg x) ,
\label{eq:g}\\
x\PQu_v(x) &= A_{\PQu_v}x^{B_{\PQu_v}}\,(1-x)^{C_{\PQu_v}}\,(1+D_{\PQu_v}x+E_{\PQu_v}x^2) ,
\label{eq:uv}\\
x\PQd_v(x) &= A_{\PQd_v}x^{B_{\PQd_v}}\,(1-x)^{C_{\PQd_v}},
\label{eq:dv}\\
x\overline{U}(x)&= A_{\overline{U}}x^{B_{\overline{U}}}\, (1-x)^{C_{\overline{U}}}\, (1+E_{\overline{U}}x^2),
\label{eq:Ubar}\\
x\overline{D}(x)&= A_{\overline{D}}x^{B_{\overline{D}}}\, (1-x)^{C_{\overline{D}}}.
\label{eq:Dbar}
\end{align}

The normalization parameters $A_{\PQu_v}$, $A_{\PQd_v}$, and $A_\cPg$ are determined by the QCD sum
rules, the $B$ parameters are responsible for the small-$x$ behavior of the PDFs, and the $C$ parameters describe the shape of
the distribution as $x \to 1$. Additional constraints $B_{\overline{U}} = B_{\overline{D}}$ and
$A_{\overline{U}} = A_{\overline{D}}(1 - f_{\PQs})$ are imposed, with $f_{\PQs}$ being the strangeness
fraction, $\PAQs/( \PAQd + \PAQs)$, which is set to $0.31\pm0.08$ as in Ref.~\cite{Martin:2009ad},
consistent with the value obtained using the CMS measurements of $\PW$+c production~\cite{Chatrchyan:2013mza}.
Using the measured values for \stt allows the addition of a new free parameter, $D_{\PQu_v}$, in Eq.~(\ref{eq:uv}), as compared to
the ana\-ly\-sis in Ref.~\cite{Khachatryan:2016pev}.

The predicted and measured cross sections for all the data sets, together with their corresponding uncertainties, are used to build a global $\chi^2$,
minimized to determine the PDF parameters~\cite{Alekhin:2014irh, herafitter}. The results of the fit are given in Table~\ref{chi2_pdffit_table}.
The quality of the overall fit can be judged based on the global $\chi^2$ divided by the number of degrees of freedom, $n_{\mathrm{dof}}$.
For each data set included in the fit, the partial $\chi^2$ divided by the
number of the measurements (data points), $n_{\mathrm{dp}}$, is also provided. The correlated part of $\chi^2$, also given in Table~\ref{chi2_pdffit_table},
quantifies the influence of the correlated systematic uncertainties in the fit.
The global and partial $\chi^2$ values indicate a general agreement among all the data sets.
The somewhat high $\chi^2/n_{\mathrm{dp}}$ values for the combined DIS data are very similar to those observed in
Ref.~\cite{Abramowicz:2015mha}, where they are investigated in detail.

\begin{table}[!tb]
\centering
\renewcommand{\arraystretch}{1.25}
\topcaption{Partial $\chi^2$ per number of data points, $n_{\mathrm{dp}}$, and the global $\chi^2$ per
degrees of freedom, $n_{\text{dof}}$, as obtained in the QCD analysis of DIS data, the CMS muon charge asymmetry measurements, and the \stt results at $\sqrts = 5.02$\TeV from this analysis.
For the HERA measurements, the energy of the proton beam ($E_{\Pp}$) is listed for each data set, with the electron/positron energy of
27.5\GeV. The correlated part of the global $\chi^2$ value is also given.}

\begin{tabular}[!tb]{l  c}
Data sets   & Partial $\chi^2/n_{\mathrm{dp}}$ \\
\hline
HERA neutral current,  $\Pep \Pp$,  $E_{\Pp}=920\GeV$ &  $449/377$  \\
HERA neutral current,  $\Pep \Pp$,  $E_{\Pp}=820\GeV$ &  $71/70$ \\
HERA neutral current,  $\Pep \Pp$,  $E_{\Pp}=575\GeV$ &  $224/254$ \\
HERA neutral current,  $\Pep \Pp$,  $E_{\Pp}=460\GeV$ &  $218/204$ \\
HERA neutral current,  $\Pem \Pp$,  $E_{\Pp}=920\GeV$ &  $218/159$ \\
HERA charged current,  $\Pep \Pp$,  $E_{\Pp}=920\GeV$ &  $43/39 $ \\
HERA charged current,  $\Pem \Pp$,  $E_{\Pp}=920\GeV$ &  $53/42 $ \\
CMS $\PW^\pm$ muon charge asymmetry  &  $2.4/11$ \\
CMS \stt, \empm, 5.02\TeV & $1.03 /1$ \\
CMS \stt, \mmpm, 5.02\TeV & $0.01 /1$ \\
CMS \stt, $\ell$+jets, 5.02\TeV & $0.70 /1$ \\[\cmsTabSkip]
Correlated $\chi^2$                                 & 100 \\
Global $\chi^2/n_{\text{dof}}$                         & $1387/1145$\\
\end{tabular}
\label{chi2_pdffit_table}
\end{table}

The experimental uncertainties in the measurements are propagated to the extracted QCD fit parameters using
the MC method~\cite{Giele:1998gw, Giele:2001mr}. In this method, 400 replicas of pseudo-data are
generated, with measured values for \stt allowed to vary within the statistical and systematic uncertainties. For each of them, the
PDF fit is performed and the uncertainty is estimated as the RMS around the central value. In Fig.~\ref{PDFs_100kGeV},
the ratio and the relative uncertainties in the gluon distributions, as obtained in the
QCD analyses with and without the measured values for \stt, are shown. A moderate reduction of the uncertainty in the
gluon distribution at $x \gtrsim 0.1$ is observed, once the measured values for \stt are included in the fit.
The uncertainties in the valence quark distributions remain unaffected. All changes in the central values of the PDFs are well
within the fit uncertainties.

Possible effects from varying the model input parameters and the initial PDF parametrization are
investigated in the same way as in the similar analysis of Ref.~\cite{Khachatryan:2016pev}.
The two cases when the measured values for \stt are included or excluded from the fit are considered, resulting in the same associated model
and parametrization uncertainties.

In conclusion, the \stt measurements at $\sqrts = 5.02$\TeV provide improved uncertainties in the gluon PDF at high $x$, though the impact is
small, owing to the large experimental uncertainties.
\begin{figure}[!htb]
\centering
\includegraphics[width=0.6\textwidth]{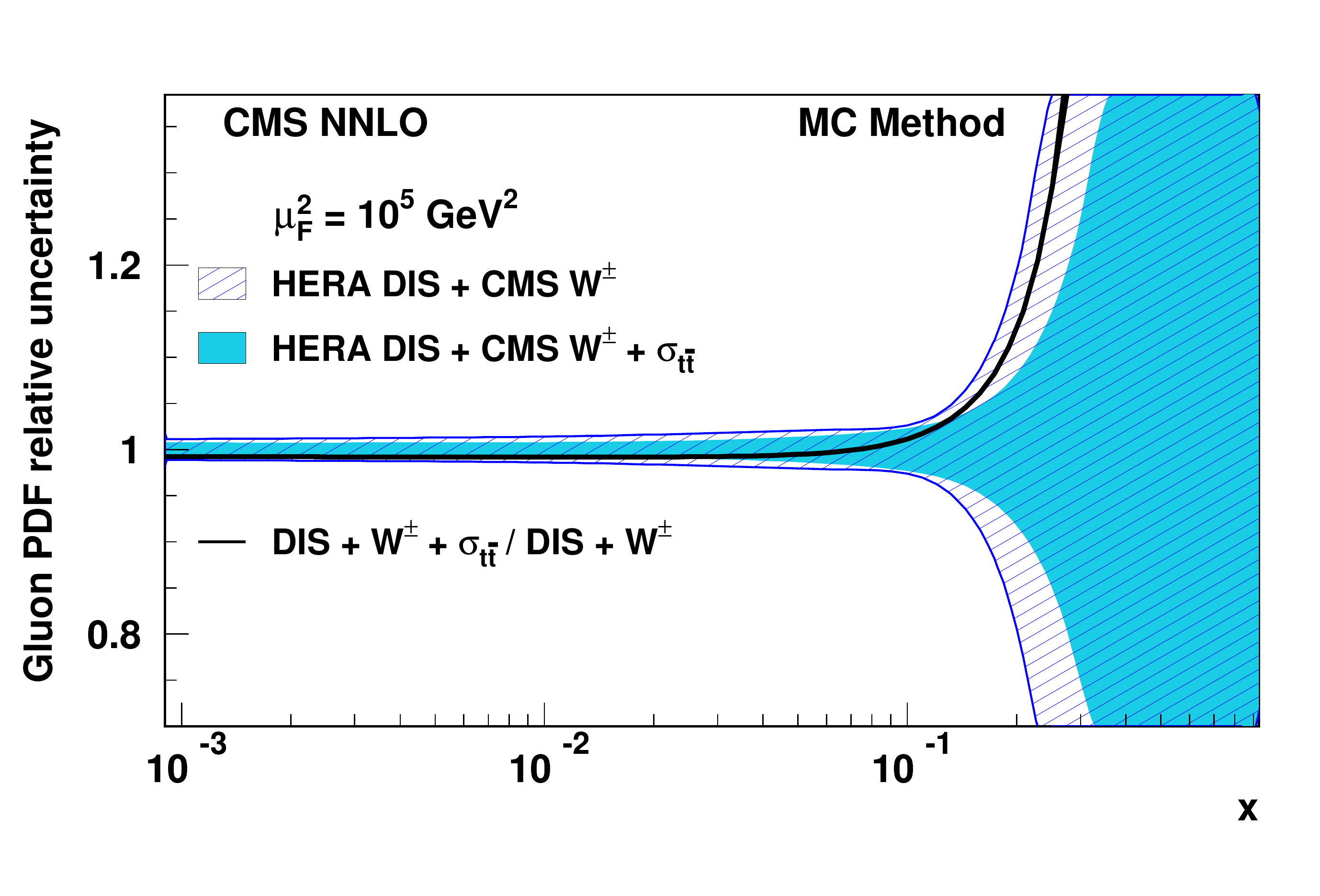}
\caption{
The relative uncertainties in the gluon distribution function of the proton as a function of $x$ at $\mu^2_{\mathrm{F}}=10^5\GeV^2$ from a
QCD analysis using the HERA DIS and CMS muon charge asymmetry measurements (hatched area), and also including the CMS \stt results
at $\sqrts = 5.02$\TeV (solid area).
The relative uncertainties are found after the two gluon distributions have been normalized to unity.
The solid line shows the ratio of the gluon distribution function found from the fit with the CMS \stt measurements included to that found without.}
\label{PDFs_100kGeV}
\end{figure}
\section{Summary}
\label{sec:summ}
The first measurement of the top quark pair (\ttbar) production cross section in pp collisions
at $\sqrts = 5.02$\TeV is presented for events with one or two leptons and at least two jets, using a data
sample collected by the CMS experiment, corresponding to an integrated luminosity of $27.4 \pm 0.6\pbinv$. The final measurement is obtained from
the combination of the measurements in the individual channels. The result is $\stt  =  69.5 \pm 6.1\stat \pm 5.6\syst \pm 1.6\lum\unit{pb}$, with a total relative uncertainty of 12\%, which is consistent with the standard model prediction.
The impact of the measured \ttbar cross section in the determination of the parton distribution functions of the proton is studied in a quantum chromodynamics
analysis at next-to-next-to-leading order. A moderate decrease of the uncertainty in the gluon distribution is observed at high values of $x$, the fractional momentum of the proton carried  by the gluon.

\begin{acknowledgments}
We congratulate our colleagues in the CERN accelerator departments for the excellent performance of the LHC and thank the technical and administrative staffs at CERN and at other CMS institutes for their contributions to the success of the CMS effort. In addition, we gratefully acknowledge the computing centers and personnel of the Worldwide LHC Computing Grid for delivering so effectively the computing infrastructure essential to our analyses. Finally, we acknowledge the enduring support for the construction and operation of the LHC and the CMS detector provided by the following funding agencies: BMWFW and FWF (Austria); FNRS and FWO (Belgium); CNPq, CAPES, FAPERJ, and FAPESP (Brazil); MES (Bulgaria); CERN; CAS, MoST, and NSFC (China); COLCIENCIAS (Colombia); MSES and CSF (Croatia); RPF (Cyprus); SENESCYT (Ecuador); MoER, ERC IUT, and ERDF (Estonia); Academy of Finland, MEC, and HIP (Finland); CEA and CNRS/IN2P3 (France); BMBF, DFG, and HGF (Germany); GSRT (Greece); OTKA and NIH (Hungary); DAE and DST (India); IPM (Iran); SFI (Ireland); INFN (Italy); MSIP and NRF (Republic of Korea); LAS (Lithuania); MOE and UM (Malaysia); BUAP, CINVESTAV, CONACYT, LNS, SEP, and UASLP-FAI (Mexico); MBIE (New Zealand); PAEC (Pakistan); MSHE and NSC (Poland); FCT (Portugal); JINR (Dubna); MON, RosAtom, RAS, RFBR and RAEP (Russia); MESTD (Serbia); SEIDI, CPAN, PCTI and FEDER (Spain); Swiss Funding Agencies (Switzerland); MST (Taipei); ThEPCenter, IPST, STAR, and NSTDA (Thailand); TUBITAK and TAEK (Turkey); NASU and SFFR (Ukraine); STFC (United Kingdom); DOE and NSF (USA).

\hyphenation{Rachada-pisek} Individuals have received support from the Marie-Curie program and the European Research Council and Horizon 2020 Grant, contract No. 675440 (European Union); the Leventis Foundation; the A. P. Sloan Foundation; the Alexander von Humboldt Foundation; the Belgian Federal Science Policy Office; the Fonds pour la Formation \`a la Recherche dans l'Industrie et dans l'Agriculture (FRIA-Belgium); the Agentschap voor Innovatie door Wetenschap en Technologie (IWT-Belgium); the Ministry of Education, Youth and Sports (MEYS) of the Czech Republic; the Council of Science and Industrial Research, India; the HOMING PLUS program of the Foundation for Polish Science, cofinanced from European Union, Regional Development Fund, the Mobility Plus program of the Ministry of Science and Higher Education, the National Science Center (Poland), contracts Harmonia 2014/14/M/ST2/00428, Opus 2014/13/B/ST2/02543, 2014/15/B/ST2/03998, and 2015/19/B/ST2/02861, Sonata-bis 2012/07/E/ST2/01406; the National Priorities Research Program by Qatar National Research Fund; the Programa Severo Ochoa del Principado de Asturias; the Thalis and Aristeia programs cofinanced by EU-ESF and the Greek NSRF; the Rachadapisek Sompot Fund for Postdoctoral Fellowship, Chulalongkorn University and the Chulalongkorn Academic into Its 2nd Century Project Advancement Project (Thailand); the Welch Foundation, contract C-1845; and the Weston Havens Foundation (USA).
\end{acknowledgments}
\bibliography{auto_generated}

\providecommand{\href}[2]{#2}\begingroup\raggedright\begin{thebibliography}{10}%
\makeatletter
\providecommand{\hrefCMSnoop }[0]{\@secondoftwo}%
\makeatother
\providecommand{\doi}{\texttt{doi:}\begingroup \urlstyle{tt}\Url}

\bibitem{polemass}
\hrefCMSnoop {}{{CMS Collaboration}, ``Determination of the top-quark pole mass
  and strong coupling constant from the $\mathrm{t}\overline{\mathrm{t}}$
  production cross section in pp collisions at {$\sqrt{s} = 7\TeV$}'',}
  \textit{ Phys. Lett. B} \textbf{ 728} (2013) 496,
  \href{http://dx.doi.org/10.1016/j.physletb.2013.12.009}{\doi{10.1016/j.physletb.2013.12.009}},
  \href{http://www.arXiv.org/abs/1307.1907}{\texttt{arXiv:1307.1907}}.
  [Erratum: \DOI{10.1016/j.physletb.2014.08.040}].

\bibitem{Sirunyan:2017azo}
\hrefCMSnoop {}{{CMS Collaboration}, ``{Measurement of double-differential
  cross sections for top quark pair production in pp collisions at $\sqrt{s} =
  8$ $\,\text {TeV}$ and impact on parton distribution functions}'',} \textit{
  Eur. Phys. J. C} \textbf{ 77} (2017) 459,
  \href{http://dx.doi.org/10.1140/epjc/s10052-017-4984-5}{\doi{10.1140/epjc/s10052-017-4984-5}},
\href{http://www.arXiv.org/abs/1703.01630}{\texttt{arXiv:1703.01630}}.

\bibitem{ATLAStt12}
\hrefCMSnoop {}{{ATLAS Collaboration}, ``{Measurement of the top quark pair
  production cross-section with ATLAS in the single lepton channel}'',}
  \textit{ Phys. Lett. B} \textbf{ 711} (2012) 244,
  \href{http://dx.doi.org/10.1016/j.physletb.2012.03.083}{\doi{10.1016/j.physletb.2012.03.083}},
\href{http://www.arXiv.org/abs/1201.1889}{\texttt{arXiv:1201.1889}}.

\bibitem{ATLAStt6}
\hrefCMSnoop {}{{ATLAS Collaboration}, ``{Measurement of the $\mathrm{t
  \bar{t}}$ production cross-section using $\mathrm{e} \mu$ events with
  $\rm{b}$-tagged jets in pp collisions at $\sqrt{s}=7$ and 8 TeV with the
  ATLAS detector}'',} \textit{ Eur. Phys. J. C} \textbf{ 74} (2014) 3109,
  \href{http://dx.doi.org/10.1140/epjc/s10052-014-3109-7}{\doi{10.1140/epjc/s10052-014-3109-7}},
\href{http://www.arXiv.org/abs/1406.5375}{\texttt{arXiv:1406.5375}}.

\bibitem{Khachatryan:2016mqs}
\hrefCMSnoop {}{{CMS Collaboration}, ``{Measurement of the \ttbar production
  cross section in the e$\mu$ channel in proton-proton collisions at $\sqrt{s}
  = 7$ and $8$ TeV}'',} \textit{ JHEP} \textbf{ 08} (2016) 029,
  \href{http://dx.doi.org/10.1007/JHEP08(2016)029}{\doi{10.1007/JHEP08(2016)029}},
\href{http://www.arXiv.org/abs/1603.02303}{\texttt{arXiv:1603.02303}}.

\bibitem{Khachatryan:2016yzq}
\hrefCMSnoop {}{{CMS Collaboration}, ``{Measurements of the \ttbar production
  cross section in lepton+jets final states in pp collisions at 8 TeV and ratio
  of 8 to 7 TeV cross sections}'',} \textit{ Eur. Phys. J. C} \textbf{ 77}
  (2017) 15,
  \href{http://dx.doi.org/10.1140/epjc/s10052-016-4504-z}{\doi{10.1140/epjc/s10052-016-4504-z}},
\href{http://www.arXiv.org/abs/1602.09024}{\texttt{arXiv:1602.09024}}.

\bibitem{ATLAStt13TeV2}
\hrefCMSnoop {}{{ATLAS Collaboration}, ``{Measurement of the $\mathrm{t
  \bar{t}}$ production cross-section using $\mathrm{e} \mu$ events with
  b-tagged jets in pp collisions at $\sqrt{s}$ = 13 TeV with the ATLAS
  detector}'',} \textit{ Phys. Lett. B} \textbf{ 761} (2016) 136,
  \href{http://dx.doi.org/10.1016/j.physletb.2016.08.019}{\doi{10.1016/j.physletb.2016.08.019}},
\href{http://www.arXiv.org/abs/1606.02699}{\texttt{arXiv:1606.02699}}.

\bibitem{Khachatryan:2015uqb}
\hrefCMSnoop {}{{CMS Collaboration}, ``{Measurement of the top quark pair
  production cross section in proton-proton collisions at $\sqrt{s} =$ 13
  TeV}'',} \textit{ Phys. Rev. Lett.} \textbf{ 116} (2016) 052002,
  \href{http://dx.doi.org/10.1103/PhysRevLett.116.052002}{\doi{10.1103/PhysRevLett.116.052002}},
\href{http://www.arXiv.org/abs/1510.05302}{\texttt{arXiv:1510.05302}}.

\bibitem{CMS-PAS-TOP-16-005}
\hrefCMSnoop {}{{CMS Collaboration}, ``{Measurement of the $\mathrm{t \bar{t}}$
  production cross section using events in the $\mathrm{e} \mu$ final state in
  pp collisions at $\sqrt{s} = $ 13 TeV}'',} \textit{ Eur. Phys. J. C} \textbf{
  77} (2017) 172,
  \href{http://dx.doi.org/10.1140/epjc/s10052-017-4718-8}{\doi{10.1140/epjc/s10052-017-4718-8}},
\href{http://www.arXiv.org/abs/1611.04040}{\texttt{arXiv:1611.04040}}.

\bibitem{CMS-PAS-TOP-16-006}
\hrefCMSnoop {}{{CMS Collaboration}, ``Measurement of the
  $\mathrm{t}\overline{\mathrm{t}}$ production cross section using events with
  one lepton and at least one jet in pp collisions at {$\sqrt{s}=13$\TeV}'',}
  \textit{ JHEP} \textbf{ 09} (2017) 051,
  \href{http://dx.doi.org/10.1007/JHEP09(2017)051}{\doi{10.1007/JHEP09(2017)051}},
  \href{http://www.arXiv.org/abs/1701.06228}{\texttt{arXiv:1701.06228}}.

\bibitem{powheg}
\hrefCMSnoop {}{S.~Frixione, P.~Nason, and C.~Oleari, ``{Matching NLO QCD
  computations with parton shower simulations: the POWHEG method}'',} \textit{
  JHEP} \textbf{ 11} (2007) 070,
  \href{http://dx.doi.org/10.1088/1126-6708/2007/11/070}{\doi{10.1088/1126-6708/2007/11/070}},
  \href{http://www.arXiv.org/abs/0709.2092}{\texttt{arXiv:0709.2092}}.

\bibitem{powheg2}
\hrefCMSnoop {}{S.~Alioli, P.~Nason, C.~Oleari, and E.~Re, ``{A general
  framework for implementing NLO calculations in shower Monte Carlo programs:
  the POWHEG BOX}'',} \textit{ JHEP} \textbf{ 06} (2010) 043,
  \href{http://dx.doi.org/10.1007/JHEP06(2010)043}{\doi{10.1007/JHEP06(2010)043}},
\href{http://www.arXiv.org/abs/1002.2581}{\texttt{arXiv:1002.2581}}.

\bibitem{powheghvq}
\hrefCMSnoop {}{S.~Frixione, P.~Nason, and G.~Ridolfi, ``{A positive-weight
  next-to-leading-order Monte Carlo for heavy flavour hadroproduction}'',}
  \textit{ JHEP} \textbf{ 0709} (2007) 126,
  \href{http://dx.doi.org/10.1088/1126-6708/2007/09/126}{\doi{10.1088/1126-6708/2007/09/126}}.

\bibitem{Ball:2014uwa}
\hrefCMSnoop {}{{NNPDF} Collaboration, ``Parton distributions for the {LHC Run
  II}'',} \textit{ JHEP} \textbf{ 04} (2015) 040,
  \href{http://dx.doi.org/10.1007/JHEP04(2015)040}{\doi{10.1007/JHEP04(2015)040}},
\href{http://www.arXiv.org/abs/1410.8849}{\texttt{arXiv:1410.8849}}.

\bibitem{dEnterria:2015jna}
\hrefCMSnoop {}{D.~d'Enterria, K.~Krajczar, and H.~Paukkunen, ``{Top-quark
  production in proton-nucleus and nucleus-nucleus collisions at LHC energies
  and beyond}'',} \textit{ Phys. Lett. B} \textbf{ 746} (2015) 64,
  \href{http://dx.doi.org/10.1016/j.physletb.2015.04.044}{\doi{10.1016/j.physletb.2015.04.044}},
\href{http://www.arXiv.org/abs/1501.05879}{\texttt{arXiv:1501.05879}}.

\bibitem{CMS-PAS-FTR-13-025}
\href {http://cds.cern.ch/record/1610789}{{CMS Collaboration}, ``{Projections
  for heavy ions with HL-LHC}'',} CMS Physics Analysis Summary
  CMS-PAS-FTR-13-025, 2013.

\bibitem{ttbarpPb}
\hrefCMSnoop {}{{CMS Collaboration}, ``Observation of top quark production in
  proton-nucleus collisions'',} \textit{ Phys. Rev. Lett.} \textbf{ 119} (2017)
  242001,
  \href{http://dx.doi.org/10.1103/PhysRevLett.119.242001}{\doi{10.1103/PhysRevLett.119.242001}},
  \href{http://www.arXiv.org/abs/1709.07411}{\texttt{arXiv:1709.07411}}.

\bibitem{hlt}
\hrefCMSnoop {}{{CMS Collaboration}, ``{The CMS trigger system}'',} \textit{
  JINST} \textbf{ 12} (2017) P01020,
  \href{http://dx.doi.org/10.1088/1748-0221/12/01/P01020}{\doi{10.1088/1748-0221/12/01/P01020}},
\href{http://www.arXiv.org/abs/1609.02366}{\texttt{arXiv:1609.02366}}.

\bibitem{Chatrchyan:2008zzk}
\hrefCMSnoop {}{{CMS Collaboration}, ``The {CMS} experiment at the {CERN}
  {LHC}'',} \textit{ JINST} \textbf{ 3} (2008) S08004,
\href{http://dx.doi.org/10.1088/1748-0221/3/08/S08004}{\doi{10.1088/1748-0221/3/08/S08004}}.

\bibitem{CMS-PAS-LUM-16-001}
\href {http://cds.cern.ch/record/2235781}{{CMS Collaboration}, ``{CMS
  luminosity calibration for the pp reference run at
  $\sqrt{s}=5.02~\mathrm{TeV}$}'',} CMS Physics Analysis Summary
  CMS-PAS-LUM-16-001, 2016.

\bibitem{Sjostrand:2006za}
\hrefCMSnoop {}{T.~Sj{\"o}strand, S.~Mrenna, and P.~Skands, ``{\PYTHIA 6.4
  physics and manual}'',} \textit{ JHEP} \textbf{ 05} (2006) 026,
  \href{http://dx.doi.org/10.1088/1126-6708/2006/05/026}{\doi{10.1088/1126-6708/2006/05/026}},
\href{http://www.arXiv.org/abs/hep-ph/0603175}{\texttt{arXiv:hep-ph/0603175}}.

\bibitem{Sjostrand:2014zea}
T.~Sj{\"o}strand\hrefCMSnoop {}{ {et~al.}, ``{An introduction to \PYTHIA
  8.2}'',} \textit{ Comput. Phys. Commun.} \textbf{ 191} (2015) 159,
  \href{http://dx.doi.org/10.1016/j.cpc.2015.01.024}{\doi{10.1016/j.cpc.2015.01.024}},
\href{http://www.arXiv.org/abs/1410.3012}{\texttt{arXiv:1410.3012}}.

\bibitem{Khachatryan:2015pea}
\hrefCMSnoop {}{{CMS Collaboration}, ``{Event generator tunes obtained from
  underlying event and multiparton scattering measurements}'',} \textit{ Eur.
  Phys. J. C} \textbf{ 76} (2016) 155,
  \href{http://dx.doi.org/10.1140/epjc/s10052-016-3988-x}{\doi{10.1140/epjc/s10052-016-3988-x}},
\href{http://www.arXiv.org/abs/1512.00815}{\texttt{arXiv:1512.00815}}.

\bibitem{Skands:2014pea}
\hrefCMSnoop {}{P.~Skands, S.~Carrazza, and J.~Rojo, ``{Tuning PYTHIA 8.1: the
  Monash 2013 tune}'',} \textit{ Eur. Phys. J. C} \textbf{ 74} (2014) 3024,
  \href{http://dx.doi.org/10.1140/epjc/s10052-014-3024-y}{\doi{10.1140/epjc/s10052-014-3024-y}},
\href{http://www.arXiv.org/abs/1404.5630}{\texttt{arXiv:1404.5630}}.

\bibitem{amcatnlo}
J.~Alwall\hrefCMSnoop {}{ {et~al.}, ``{The automated computation of tree-level
  and next-to-leading order differential cross sections, and their matching to
  parton shower simulations}'',} \textit{ JHEP} \textbf{ 07} (2014) 079,
  \href{http://dx.doi.org/10.1007/JHEP07(2014)079}{\doi{10.1007/JHEP07(2014)079}},
\href{http://www.arXiv.org/abs/1405.0301}{\texttt{arXiv:1405.0301}}.

\bibitem{fxfx}
\hrefCMSnoop {}{R.~Frederix and S.~Frixione, ``{Merging meets matching in
  MC@NLO}'',} \textit{ JHEP} \textbf{ 12} (2012) 061,
  \href{http://dx.doi.org/10.1007/JHEP12(2012)061}{\doi{10.1007/JHEP12(2012)061}},
\href{http://www.arXiv.org/abs/1209.6215}{\texttt{arXiv:1209.6215}}.

\bibitem{fewz}
\hrefCMSnoop {}{K.~Melnikov and F.~Petriello, ``{Electroweak gauge boson
  production at hadron colliders through $O(\alpha_s^2)$}'',} \textit{ Phys.
  Rev. D} \textbf{ 74} (2006) 114017,
  \href{http://dx.doi.org/10.1103/PhysRevD.74.114017}{\doi{10.1103/PhysRevD.74.114017}},
\href{http://www.arXiv.org/abs/hep-ph/0609070}{\texttt{arXiv:hep-ph/0609070}}.

\bibitem{powheg1}
\hrefCMSnoop {}{S.~Alioli, P.~Nason, C.~Oleari, and E.~Re, ``{NLO single-top
  production matched with shower in POWHEG: $s$- and $t$-channel
  contributions}'',} \textit{ JHEP} \textbf{ 09} (2009) 111,
  \href{http://dx.doi.org/10.1088/1126-6708/2009/09/111}{\doi{10.1088/1126-6708/2009/09/111}},
  \href{http://www.arXiv.org/abs/0907.4076}{\texttt{arXiv:0907.4076}}.
[Erratum: \DOI{10.1007/JHEP02(2010)011}].

\bibitem{powheg3}
\hrefCMSnoop {}{E.~Re, ``Single-top {$Wt$-channel} production matched with
  parton showers using the {POWHEG} method'',} \textit{ Eur. Phys. J. C}
  \textbf{ 71} (2011) 1547,
  \href{http://dx.doi.org/10.1140/epjc/s10052-011-1547-z}{\doi{10.1140/epjc/s10052-011-1547-z}},
\href{http://www.arXiv.org/abs/1009.2450}{\texttt{arXiv:1009.2450}}.

\bibitem{Kidonakis:2013zqa}
\hrefCMSnoop {}{N.~Kidonakis, ``{Top quark production}'',} in \textit{
  {Proceedings, Helmholtz International Summer School on Physics of Heavy
  Quarks and Hadrons (HQ 2013)}}, p.~139.
\newblock Verlag Deutsches Elektronen-Synchrotron, Hamburg, 2014.
\newblock \href{http://www.arXiv.org/abs/1311.0283}{\texttt{arXiv:1311.0283}}.
\newblock
\href{http://dx.doi.org/10.3204/DESY-PROC-2013-03/Kidonakis}{\doi{10.3204/DESY-PROC-2013-03/Kidonakis}}.

\bibitem{mcfm}
\hrefCMSnoop {}{J.~M. Campbell and R.~K. Ellis, ``{\MCFM for the Tevatron and
  the LHC}'',} \textit{ Nucl. Phys. Proc. Suppl.} \textbf{ 205--206} (2010) 10,
  \href{http://dx.doi.org/10.1016/j.nuclphysbps.2010.08.011}{\doi{10.1016/j.nuclphysbps.2010.08.011}},
  \href{http://www.arXiv.org/abs/1007.3492}{\texttt{arXiv:1007.3492}}.

\bibitem{geant}
\hrefCMSnoop {}{{\textsc{Geant}4} Collaboration, ``\textsc{Geant}4 --- a
  simulation toolkit'',} \textit{ Nucl. Instrum. Meth. A} \textbf{ 506} (2003)
  250,
\href{http://dx.doi.org/10.1016/S0168-9002(03)01368-8}{\doi{10.1016/S0168-9002(03)01368-8}}.

\bibitem{top++}
\hrefCMSnoop {}{M.~Czakon and A.~Mitov, ``{\textsc{Top++}: a program for the
  calculation of the top-pair cross-section at hadron colliders}'',} \textit{
  Comput. Phys. Commun.} \textbf{ 185} (2014) 2930,
  \href{http://dx.doi.org/10.1016/j.cpc.2014.06.021}{\doi{10.1016/j.cpc.2014.06.021}},
\href{http://www.arXiv.org/abs/1112.5675}{\texttt{arXiv:1112.5675}}.

\bibitem{mitov}
\hrefCMSnoop {}{{M. Czakon, P. Fiedler and A. Mitov}, ``{Total top-quark
  pair-production cross section at hadron colliders through
  O($\alpha_S^4$)}'',} \textit{ Phys. Rev. Lett.} \textbf{ 110} (2013) 252004,
  \href{http://dx.doi.org/10.1103/PhysRevLett.110.252004}{\doi{10.1103/PhysRevLett.110.252004}},
\href{http://www.arXiv.org/abs/1303.6254}{\texttt{arXiv:1303.6254}}.

\bibitem{Wenninger:2254678}
\hrefCMSnoop {}{E.~Todesco and J.~Wenninger, ``Large hadron collider momentum
  calibration and accuracy'',} \textit{ Phys. Rev. Accel. Beams} \textbf{ 20}
  (2017) 081003,
  \href{http://dx.doi.org/10.1103/PhysRevAccelBeams.20.081003}{\doi{10.1103/PhysRevAccelBeams.20.081003}}.

\bibitem{PFPAS1}
\hrefCMSnoop {}{{CMS Collaboration}, ``Particle-flow reconstruction and global
  event description with the cms detector'',} \textit{ JINST} \textbf{ 12}
  (2017) P10003,
  \href{http://dx.doi.org/10.1088/1748-0221/12/10/P10003}{\doi{10.1088/1748-0221/12/10/P10003}},
\href{http://www.arXiv.org/abs/1706.04965}{\texttt{arXiv:1706.04965}}.

\bibitem{Khachatryan:2015hwa}
\hrefCMSnoop {}{{CMS Collaboration}, ``{Performance of electron reconstruction
  and selection with the CMS detector in proton-proton collisions at $\sqrt{s}
  = 8$\TeV}'',} \textit{ JINST} \textbf{ 10} (2015) P06005,
  \href{http://dx.doi.org/10.1088/1748-0221/10/06/P06005}{\doi{10.1088/1748-0221/10/06/P06005}},
\href{http://www.arXiv.org/abs/1502.02701}{\texttt{arXiv:1502.02701}}.

\bibitem{Chatrchyan:2012xi}
\hrefCMSnoop {}{{CMS Collaboration}, ``{Performance of CMS muon reconstruction
  in pp collision events at $\sqrt{s} = 7$\TeV}'',} \textit{ JINST} \textbf{ 7}
  (2012) P10002,
  \href{http://dx.doi.org/10.1088/1748-0221/7/10/P10002}{\doi{10.1088/1748-0221/7/10/P10002}},
\href{http://www.arXiv.org/abs/1206.4071}{\texttt{arXiv:1206.4071}}.

\bibitem{muid}
\hrefCMSnoop {}{{CMS Collaboration}, ``{The performance of the CMS muon
  detector in proton-proton collisions at $\sqrt{s} = 7\TeV$ at the LHC}'',}
  \textit{ JINST} \textbf{ 8} (2013) P11002,
  \href{http://dx.doi.org/10.1088/1748-0221/8/11/P11002}{\doi{10.1088/1748-0221/8/11/P11002}},
\href{http://www.arXiv.org/abs/1306.6905}{\texttt{arXiv:1306.6905}}.

\bibitem{CMS-PAS-JME-16-004}
\href {https://cds.cern.ch/record/2205284}{{CMS Collaboration}, ``{Performance
  of missing energy reconstruction in $\sqrt{s}=13~\mathrm{TeV}$ pp collision
  data using the CMS detector}'',} CMS Physics Analysis Summary
  CMS-PAS-JME-16-004, 2016.

\bibitem{antikt}
\hrefCMSnoop {}{M.~Cacciari, G.~P. Salam, and G.~Soyez, ``The anti-$k_{t}$ jet
  clustering algorithm'',} \textit{ JHEP} \textbf{ 04} (2008) 063,
  \href{http://dx.doi.org/10.1088/1126-6708/2008/04/063}{\doi{10.1088/1126-6708/2008/04/063}},
\href{http://www.arXiv.org/abs/0802.1189}{\texttt{arXiv:0802.1189}}.

\bibitem{Cacciari:2011ma}
\hrefCMSnoop {}{M.~Cacciari, G.~P. Salam, and G.~Soyez, ``{FastJet} user
  manual'',} \textit{ Eur. Phys. J. C} \textbf{ 72} (2012) 1896,
  \href{http://dx.doi.org/10.1140/epjc/s10052-012-1896-2}{\doi{10.1140/epjc/s10052-012-1896-2}},
\href{http://www.arXiv.org/abs/1111.6097}{\texttt{arXiv:1111.6097}}.

\bibitem{Cacciari:2007fd}
\hrefCMSnoop {}{M.~Cacciari and G.~P. Salam, ``{Pileup subtraction using jet
  areas}'',} \textit{ Phys. Lett. B} \textbf{ 659} (2008) 119,
  \href{http://dx.doi.org/10.1016/j.physletb.2007.09.077}{\doi{10.1016/j.physletb.2007.09.077}},
\href{http://www.arXiv.org/abs/0707.1378}{\texttt{arXiv:0707.1378}}.

\bibitem{inclusWZ3pb}
\hrefCMSnoop {}{{CMS Collaboration}, ``{Measurements of inclusive W and Z cross
  sections in pp collisions at $\sqrt{s} = 7\TeV$}'',} \textit{ JHEP} \textbf{
  01} (2011) 080,
  \href{http://dx.doi.org/10.1007/JHEP01(2011)080}{\doi{10.1007/JHEP01(2011)080}},
  \href{http://www.arXiv.org/abs/1012.2466}{\texttt{arXiv:1012.2466}}.

\bibitem{JESPUB}
\hrefCMSnoop {}{{CMS Collaboration}, ``Determination of jet energy calibration
  and transverse momentum resolution in {CMS}'',} \textit{ JINST} \textbf{ 6}
  (2011) P11002,
  \href{http://dx.doi.org/10.1088/1748-0221/6/11/P11002}{\doi{10.1088/1748-0221/6/11/P11002}},
  \href{http://www.arXiv.org/abs/1107.4277}{\texttt{arXiv:1107.4277}}.

\bibitem{btaggingRun2}
\hrefCMSnoop {}{{CMS Collaboration}, ``Identification of heavy-flavour jets
  with the {CMS} detector in pp collisions at 13 {TeV}'',} (2017).
  \href{http://www.arXiv.org/abs/1712.07158}{\texttt{arXiv:1712.07158}}.
Submitted to \textit{JINST}.

\bibitem{Khachatryan:2010ez}
\hrefCMSnoop {}{{CMS Collaboration}, ``{First measurement of the cross section
  for top-quark pair production in proton-proton collisions at $\sqrt{s}=7$
  TeV}'',} \textit{ Phys. Lett. B} \textbf{ 695} (2011) 424,
  \href{http://dx.doi.org/10.1016/j.physletb.2010.11.058}{\doi{10.1016/j.physletb.2010.11.058}},
\href{http://www.arXiv.org/abs/1010.5994}{\texttt{arXiv:1010.5994}}.

\bibitem{JES8TeVPUB}
\hrefCMSnoop {}{{CMS Collaboration}, ``Jet energy scale and resolution in the
  cms experiment in pp collisions at 8 tev'',} \textit{ JINST} \textbf{ 12}
  (2017) P02014,
  \href{http://dx.doi.org/10.1088/1748-0221/12/02/P02014}{\doi{10.1088/1748-0221/12/02/P02014}},
  \href{http://www.arXiv.org/abs/1607.03663}{\texttt{arXiv:1607.03663}}.

\bibitem{herwigpp}
M.~B{\"a}hr\hrefCMSnoop {}{ {et~al.}, ``{\HERWIG}++ physics and manual'',}
  \textit{ Eur. Phys. J. C} \textbf{ 58} (2008) 639,
  \href{http://dx.doi.org/10.1140/epjc/s10052-008-0798-9}{\doi{10.1140/epjc/s10052-008-0798-9}},
\href{http://www.arXiv.org/abs/0803.0883}{\texttt{arXiv:0803.0883}}.

\bibitem{Cowan:2010js}
\hrefCMSnoop {}{G.~Cowan, K.~Cranmer, E.~Gross, and O.~Vitells, ``Asymptotic
  formulae for likelihood-based tests of new physics'',} \textit{ Eur. Phys. J.
  C} \textbf{ 71} (2011) 1554,
  \href{http://dx.doi.org/10.1140/epjc/s10052-011-1554-0}{\doi{10.1140/epjc/s10052-011-1554-0}},
  \href{http://www.arXiv.org/abs/1007.1727}{\texttt{arXiv:1007.1727}}.
[Erratum: \DOI{10.1140/epjc/s10052-013-2501-z}].

\bibitem{Lyons:1988rp}
\hrefCMSnoop {}{L.~Lyons, D.~Gibaut, and P.~Clifford, ``{How to combine
  correlated estimates of a single physical quantity}'',} \textit{ Nucl.
  Instrum. Meth. A} \textbf{ 270} (1988) 110,
  \href{http://dx.doi.org/10.1016/0168-9002(88)90018-6}{\doi{10.1016/0168-9002(88)90018-6}}.

\bibitem{Valassi:2013bga}
\hrefCMSnoop {}{A.~Valassi and R.~Chierici, ``{Information and treatment of
  unknown correlations in the combination of measurements using the BLUE
  method}'',} \textit{ Eur. Phys. J. C} \textbf{ 74} (2014) 2717,
  \href{http://dx.doi.org/10.1140/epjc/s10052-014-2717-6}{\doi{10.1140/epjc/s10052-014-2717-6}},
\href{http://www.arXiv.org/abs/1307.4003}{\texttt{arXiv:1307.4003}}.

\bibitem{Lista:2014qia}
\hrefCMSnoop {}{L.~Lista, ``{The bias of the unbiased estimator: a study of the
  iterative application of the BLUE method}'',} \textit{ Nucl. Instrum. Meth.
  A} \textbf{ 764} (2014) 82,
  \href{http://dx.doi.org/10.1016/j.nima.2014.07.021}{\doi{10.1016/j.nima.2014.07.021}},
  \href{http://www.arXiv.org/abs/1405.3425}{\texttt{arXiv:1405.3425}}.
[Erratum: \DOI{10.1016/j.nima.2014.11.054}].

\bibitem{Harland-Lang:2014zoa}
\hrefCMSnoop {}{L.~A. Harland-Lang, A.~D. Martin, P.~Motylinski, and R.~S.
  Thorne, ``{Parton distributions in the LHC era: MMHT 2014 PDFs}'',} \textit{
  Eur. Phys. J. C} \textbf{ 75} (2015) 204,
  \href{http://dx.doi.org/10.1140/epjc/s10052-015-3397-6}{\doi{10.1140/epjc/s10052-015-3397-6}},
\href{http://www.arXiv.org/abs/1412.3989}{\texttt{arXiv:1412.3989}}.

\bibitem{Dulat:2015mca}
S.~Dulat\hrefCMSnoop {}{ {et~al.}, ``{New parton distribution functions from a
  global analysis of quantum chromodynamics}'',} \textit{ Phys. Rev. D}
  \textbf{ 93} (2016) 033006,
  \href{http://dx.doi.org/10.1103/PhysRevD.93.033006}{\doi{10.1103/PhysRevD.93.033006}},
\href{http://www.arXiv.org/abs/1506.07443}{\texttt{arXiv:1506.07443}}.

\bibitem{Alekhin:2017kpj}
\hrefCMSnoop {}{S.~Alekhin, J.~Bl{\"u}mlein, S.~Moch, and R.~Placakyte,
  ``{Parton distribution functions, $\alpha_s$, and heavy-quark masses for LHC
  Run II}'',} \textit{ Phys. Rev. D} \textbf{ 96} (2017) 014011,
  \href{http://dx.doi.org/10.1103/PhysRevD.96.014011}{\doi{10.1103/PhysRevD.96.014011}},
\href{http://www.arXiv.org/abs/1701.05838}{\texttt{arXiv:1701.05838}}.

\bibitem{Abramowicz:2015mha}
\hrefCMSnoop {}{{ZEUS and H1 Collaborations}, ``{Combination of measurements of
  inclusive deep inelastic $\mathrm{e^{\pm }p}$ scattering cross sections and
  QCD analysis of HERA data}'',} \textit{ Eur. Phys. J. C} \textbf{ 75} (2015)
  580,
  \href{http://dx.doi.org/10.1140/epjc/s10052-015-3710-4}{\doi{10.1140/epjc/s10052-015-3710-4}},
\href{http://www.arXiv.org/abs/1506.06042}{\texttt{arXiv:1506.06042}}.

\bibitem{Khachatryan:2016pev}
\hrefCMSnoop {}{{CMS Collaboration}, ``{Measurement of the differential cross
  section and charge asymmetry for inclusive $\mathrm {p}\mathrm {p}\rightarrow
  \mathrm {W}^{\pm }+X$ production at ${\sqrt{s}} = 8$ TeV}'',} \textit{ Eur.
  Phys. J. C} \textbf{ 76} (2016) 469,
  \href{http://dx.doi.org/10.1140/epjc/s10052-016-4293-4}{\doi{10.1140/epjc/s10052-016-4293-4}},
\href{http://www.arXiv.org/abs/1603.01803}{\texttt{arXiv:1603.01803}}.

\bibitem{Alekhin:2014irh}
\hrefCMSnoop {}{S.~Alekhin {et~al.}, ``{HERAFitter, open source QCD fit
  project}'',} \textit{ Eur. Phys. J. C} \textbf{ 75} (2015) 304,
  \href{http://dx.doi.org/10.1140/epjc/s10052-015-3480-z}{\doi{10.1140/epjc/s10052-015-3480-z}},
\href{http://www.arXiv.org/abs/1410.4412}{\texttt{arXiv:1410.4412}}.

\bibitem{herafitter}
\href {www.xfitter.org}{{xFitter} Collaboration}. \url{http://www.xfitter.org}.
\newblock \url {www.xfitter.org}.

\bibitem{Gribov:1972ri}
\hrefCMSnoop {}{V.~N. Gribov and L.~N. Lipatov, ``{Deep inelastic ep scattering
  in perturbation theory}'',} \textit{ Sov. J. Nucl. Phys.} \textbf{ 15} (1972)
438.

\bibitem{Altarelli:1977zs}
\hrefCMSnoop {}{G.~Altarelli and G.~Parisi, ``{Asymptotic freedom in parton
  language}'',} \textit{ Nucl. Phys. B} \textbf{ 126} (1977) 298,
\href{http://dx.doi.org/10.1016/0550-3213(77)90384-4}{\doi{10.1016/0550-3213(77)90384-4}}.

\bibitem{Curci:1980uw}
\hrefCMSnoop {}{G.~Curci, W.~Furmanski, and R.~Petronzio, ``{Evolution of
  parton densities beyond leading order: The non-singlet case}'',} \textit{
  Nucl. Phys. B} \textbf{ 175} (1980) 27,
\href{http://dx.doi.org/10.1016/0550-3213(80)90003-6}{\doi{10.1016/0550-3213(80)90003-6}}.

\bibitem{Furmanski:1980cm}
\hrefCMSnoop {}{W.~Furmanski and R.~Petronzio, ``{Singlet parton densities
  beyond leading order}'',} \textit{ Phys. Lett. B} \textbf{ 97} (1980) 437,
\href{http://dx.doi.org/10.1016/0370-2693(80)90636-X}{\doi{10.1016/0370-2693(80)90636-X}}.

\bibitem{Moch:2004pa}
\hrefCMSnoop {}{S.~Moch, J.~A.~M. Vermaseren, and A.~Vogt, ``{The three-loop
  splitting functions in QCD: the non-singlet case}'',} \textit{ Nucl. Phys. B}
  \textbf{ 688} (2004) 101,
  \href{http://dx.doi.org/10.1016/j.nuclphysb.2004.03.030}{\doi{10.1016/j.nuclphysb.2004.03.030}},
\href{http://www.arXiv.org/abs/hep-ph/0403192}{\texttt{arXiv:hep-ph/0403192}}.

\bibitem{Vogt:2004mw}
\hrefCMSnoop {}{A.~Vogt, S.~Moch, and J.~A.~M. Vermaseren, ``{The three-loop
  splitting functions in QCD: the singlet case}'',} \textit{ Nucl. Phys. B}
  \textbf{ 691} (2004) 129,
  \href{http://dx.doi.org/10.1016/j.nuclphysb.2004.04.024}{\doi{10.1016/j.nuclphysb.2004.04.024}},
\href{http://www.arXiv.org/abs/hep-ph/0404111}{\texttt{arXiv:hep-ph/0404111}}.

\bibitem{Botje:2010ay}
\hrefCMSnoop {}{M.~Botje, ``{QCDNUM: Fast QCD evolution and convolution}'',}
  \textit{ Comput. Phys. Commun.} \textbf{ 182} (2011) 490,
  \href{http://dx.doi.org/10.1016/j.cpc.2010.10.020}{\doi{10.1016/j.cpc.2010.10.020}},
\href{http://www.arXiv.org/abs/1005.1481}{\texttt{arXiv:1005.1481}}.

\bibitem{Aliev:2010zk}
M.~Aliev\hrefCMSnoop {}{ {et~al.}, ``{HATHOR: HAdronic Top and Heavy quarks
  crOss section calculatoR}'',} \textit{ Comput. Phys. Commun.} \textbf{ 182}
  (2011) 1034,
  \href{http://dx.doi.org/10.1016/j.cpc.2010.12.040}{\doi{10.1016/j.cpc.2010.12.040}},
\href{http://www.arXiv.org/abs/1007.1327}{\texttt{arXiv:1007.1327}}.

\bibitem{Martin:2009ad}
\hrefCMSnoop {}{A.~D. Martin, W.~J. Stirling, R.~S. Thorne, and G.~Watt,
  ``{Parton distributions for the LHC}'',} \textit{ Eur. Phys. J. C} \textbf{
  63} (2009) 189,
  \href{http://dx.doi.org/10.1140/epjc/s10052-009-1072-5}{\doi{10.1140/epjc/s10052-009-1072-5}},
\href{http://www.arXiv.org/abs/0901.0002}{\texttt{arXiv:0901.0002}}.

\bibitem{Chatrchyan:2013mza}
\hrefCMSnoop {}{{CMS Collaboration}, ``{Measurement of the muon charge
  asymmetry in inclusive $\mathrm{pp \to W+X}$ production at $\sqrt{s} = 7\TeV$
  and an improved determination of light parton distribution functions}'',}
  \textit{ Phys. Rev. D} \textbf{ 90} (2014) 032004,
  \href{http://dx.doi.org/10.1103/PhysRevD.90.032004}{\doi{10.1103/PhysRevD.90.032004}},
\href{http://www.arXiv.org/abs/1312.6283}{\texttt{arXiv:1312.6283}}.

\bibitem{Giele:1998gw}
\hrefCMSnoop {}{W.~T. Giele and S.~Keller, ``{Implications of hadron collider
  observables on parton distribution function uncertainties}'',} \textit{ Phys.
  Rev. D} \textbf{ 58} (1998) 094023,
  \href{http://dx.doi.org/10.1103/PhysRevD.58.094023}{\doi{10.1103/PhysRevD.58.094023}},
\href{http://www.arXiv.org/abs/hep-ph/9803393}{\texttt{arXiv:hep-ph/9803393}}.

\bibitem{Giele:2001mr}
\hrefCMSnoop {}{W.~T. Giele, S.~A. Keller, and D.~A. Kosower, ``{Parton
  distribution function uncertainties}'',} (2001).
\href{http://www.arXiv.org/abs/hep-ph/0104052}{\texttt{arXiv:hep-ph/0104052}}.

\end{thebibliography}\endgroup

\cleardoublepage \appendix\section{The CMS Collaboration \label{app:collab}}\begin{sloppypar}\hyphenpenalty=5000\widowpenalty=500\clubpenalty=5000\textbf{Yerevan Physics Institute,  Yerevan,  Armenia}\\*[0pt]
A.M.~Sirunyan, A.~Tumasyan
\vskip\cmsinstskip
\textbf{Institut f\"{u}r Hochenergiephysik,  Wien,  Austria}\\*[0pt]
W.~Adam, F.~Ambrogi, E.~Asilar, T.~Bergauer, J.~Brandstetter, E.~Brondolin, M.~Dragicevic, J.~Er\"{o}, M.~Flechl, M.~Friedl, R.~Fr\"{u}hwirth\cmsAuthorMark{1}, V.M.~Ghete, J.~Grossmann, J.~Hrubec, M.~Jeitler\cmsAuthorMark{1}, A.~K\"{o}nig, N.~Krammer, I.~Kr\"{a}tschmer, D.~Liko, T.~Madlener, I.~Mikulec, E.~Pree, D.~Rabady, N.~Rad, H.~Rohringer, J.~Schieck\cmsAuthorMark{1}, R.~Sch\"{o}fbeck, M.~Spanring, D.~Spitzbart, W.~Waltenberger, J.~Wittmann, C.-E.~Wulz\cmsAuthorMark{1}, M.~Zarucki
% ...
\vskip\cmsinstskip
% ...
1:~~Also at Vienna University of Technology, Vienna, Austria\\
% ...

\end{sloppypar}
\end{document}